\documentclass[reprint,superscriptaddress,amsmath,amssymb,aps,floatfix,nofootinbib,longbibliography]{revtex4-1}

\usepackage[pdftex]{graphicx}
\usepackage{bm}
\usepackage{color}
\usepackage{times}
\usepackage{amsmath}
\usepackage{amssymb}

\begin{document}

\title{Dynamics of swelling and drying in a spherical gel}

\author{Thibault Bertrand}
\email{thibault.bertrand@yale.edu}
\affiliation{Department of Mechanical Engineering and Materials Science, Yale University, New Haven, Connecticut 06520, USA}
\author{Jorge Peixinho}
\affiliation{Laboratoire Ondes et Milieux Complexes, CNRS \& Normandie Universit\'e, Le Havre, 76600, France}
\author{Shomeek Mukhopadhyay} 
\affiliation{Department of Geology and Geophysics, Yale University, New Haven, Connecticut, 06511, USA}
\author{Christopher W. MacMinn}
\email{christopher.macminn@eng.ox.ac.uk}
\affiliation{Department of Engineering Science, University of Oxford, Oxford, OX1 3PJ, UK}

\date{\today}


\begin{abstract}
Swelling is a volumetric-growth process in which a porous material expands by spontaneous imbibition of additional pore fluid. Swelling is distinct from other growth processes in that it is inherently poromechanical: Local expansion of the pore structure requires that additional fluid be drawn from elsewhere in the material, or into the material from across the boundaries. Here, we study the swelling and subsequent drying of a sphere of hydrogel. We develop a dynamic model based on large-deformation poromechanics and the theory of ideal elastomeric gels, and we compare the predictions of this model with a series of experiments performed with polyacrylamide spheres. We use the model and the experiments to study the complex internal dynamics of swelling and drying, and to highlight the fundamentally transient nature of these strikingly different processes. Although we assume spherical symmetry, the model also provides insight into the transient patterns that form and then vanish during swelling as well as the risk of fracture during drying.
\end{abstract}

\maketitle

\section{Introduction}
\label{s:intro}

Swelling is a fundamental process in biology, engineering, and the earth sciences: Tissues swell after injury, wooden structures swell with humidity, and dry soils swell after rainfall. Macroscopically, swelling is the volumetric growth of a porous material due to the spontaneous imbibition of additional pore fluid. Swelling is distinct from other growth processes because of the fundamental role of hydrodynamics: Local expansion of the pore structure is coupled to the evolving fluid distribution, making swelling inherently dynamic and poromechanical.

Swelling in polymeric gels is a classical topic in soft matter, primarily from the perspective of chemical physics~\cite{flory-jcp-1943b, quesada-perez-softmatter-2011a}. The mechanics of gels have attracted great interest more recently in the context of hydrogels~\cite{tanaka-n-1987a, hong-jmps-2008a, doi-jpsjpn-2009, chester-jmps-2010a}. A hydrogel is a crosslinked network of hydrophilic polymers saturated with water. Hydrogels can experience extremely large and reversible changes in volume during swelling, which can result in complex changes in shape and the development of surface patterns~\cite{tanaka-n-1987a, dervaux-prl-2011a, barros-softmatter-2012a, tallinen-natphys-2016, takahashi-softmatter-2016}. Hydrogels have found a wide variety of practical applications; for example, they are widely used for moisture absorption and in soft contact lenses~\cite{zohuriaan-mehr-jmatsci-2010, peppas-annrevbiomedeng-2000, calo-eurpolymj-2015}. In biomedical engineering, they are used for drug delivery, wound dressing, and as a scaffold for tissue engineering~\cite{langer-nature-1998, peppas-annrevbiomedeng-2000, cohen-sjam-1988a, calo-eurpolymj-2015}. They have also shown promise for use as sensors, actuators, and flow controllers~\cite{harmon-polymer-2003, deligkaris-snb-2010, holmes-softmatter-2011}, and as a model system in soft granular matter~\cite{mukhopadhyay-pre-2011a, macminn-prx-2015a, brodu-natcomm-2015a}.

In applications involving swelling, such as moisture absorption, drug delivery, and sensing and actuation, the primary design considerations are the degree of swelling and the rate of swelling in response to various environmental stimuli. The degree of swelling is an equilibrium property of a given gel in a given environment, and is now relatively well understood~\cite{quesada-perez-softmatter-2011a, cai-epl-2012a, li-softmatter-2012a}. The rate of swelling, in contrast, is an emergent property of a gel-environment system that also depends on the gel geometry through the transient kinetics and mechanics of swelling. The ability to model and tune the rate of swelling in response to different stimuli is central to engineering design; for example, applications in actuation and flow control rely on changes in size and/or shape during swelling and are typically designed for a fast response, whereas contact lenses should tend to preserve their size and shape and should respond slowly in order to buffer the eye from sudden variations in ambient conditions. However, the transient mechanics of swelling have received comparatively little attention and remain poorly understood. For relatively small volume changes, the dynamics of swelling have been studied using both simple linear models~\cite{tanaka-jchemphys-1973, tanaka-jchemphys-1979, johnson-jchemphys-1982, scherer-jnoncrystsolids-1989, hui-jchemphys-2005, wahrmund-macromol-2009a, yoon-softmatter-2010a} and fully nonlinear models~\cite{hong-jmps-2008a, duda-jmps-2010, chester-jmps-2010a, bouklas-jmps-2015a}, but no study has yet combined the fully nonlinear and transient mechanics of swelling with the extreme volume changes that are one of the most noteworthy, surprising, and useful characteristics of hydrogels. This is due in part to the fact that transient phenomena with large volume changes and strong poromechanical coupling are very challenging from the perspective of computational mechanics.

\begin{figure*}
    \centering
    \includegraphics[width=17.2cm]{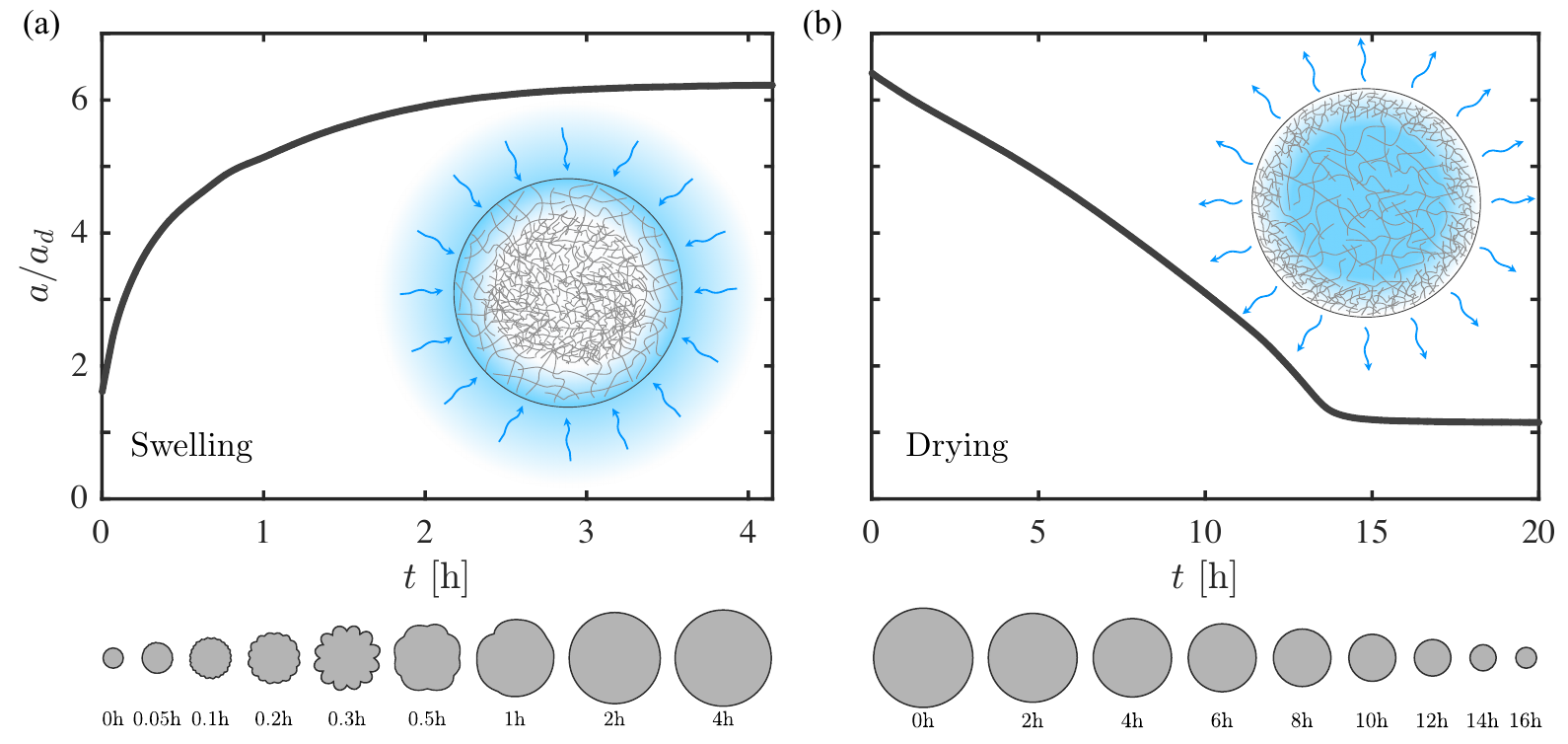}
    \caption{A polymeric hydrogel is a crosslinked network of polymer chains saturated with water. Swelling occurs due to the spontaneous imbibition of additional water, stretching the polymer chains; drying or de-swelling is the reverse. Here we show the evolution of the mean radius of beads with dry radius $a_d=0.76\,\mathrm{mm}$ and fully swollen radius $6.4a_d$ during (a)~swelling and (b)~drying with cartoons illustrating the composition. \label{fig:intro} }
\end{figure*}

Here, we focus on the simplest three-dimensional example of extreme swelling: The swelling and subsequent drying (de-swelling) of a hydrogel sphere~(Fig.~\ref{fig:intro}). Despite the apparent simplicity of this problem, no model has yet shown satisfying agreement with experiments in terms of the dynamics of swelling and drying~\cite{engelsberg-pre-2013a}. We address this problem with a fully nonlinear model that combines the framework of large-deformation poromechanics~\cite{macminn-prapplied-2016} with the theory of ideal elastomeric gels~\cite{hong-jmps-2008a, cai-epl-2012a, li-softmatter-2012a}. By including only the essential features of swelling, our approach allows for a clear and detailed exploration of the transient poromechanics of swelling and drying across a wide range of parameters; the resulting spherically symmetric model is also well-suited to efficient numerical solution, even for very large changes in volume and strongly nonlinear constitutive behavior.

For both swelling and drying, we study the full transition from one equilibrium state to another, comparing the macroscopic predictions of the model with a series of experiments. We then use the model to study the detailed mechanics of swelling and drying, highlighting the fundamental and striking differences between these two processes. We also develop a novel model for evaporation-limited drying, and we study the impact of an evaporation limit on the development of strong tensile effective stresses during drying. Although we assume spherical symmetry, the model also provides insight into the transient patterns that form and then vanish during swelling~(Figs.~\ref{fig:intro}a and \ref{fig:swelling_experiment}c), as well as the risk of fracture during drying.

The most important conclusion of our study is that swelling and drying are inherently dynamic processes. Both the development of patterns during swelling and the risk of fracture during drying are \textit{transient} phenomena that must be studied with a truly dynamic model that accounts for the evolving heterogeneous water content.

\section{Poromechanical swelling model}
\label{s:poromechanical}

A gel is a mixture of fluid and solid, where the solid forms a connected porous skeleton and the fluid occupies the pore space. In a polymeric hydrogel, the solid is a crosslinked network of polymer chains and the fluid is water. Fully swollen hydrogels typically have a solid volume fraction of less than 1\% (\textit{i.e.}, a volume swelling ratio of several hundred).

\subsection{Ideal elastomeric gels}

The swelling of a polymeric gel occurs through the spontaneous imbibition of additional fluid, which requires volumetric expansion of the polymer network to increase the pore volume. This is driven by a strong chemical affinity between the fluid and the polymer chains, such that the increase in fluid content is associated with a decrease in the free energy of the mixture. This decrease in free energy due to mixing is opposed by an increase in free energy due to elastic stretching of the polymer network. Swelling reaches equilibrium when the penalty due to further stretching precisely balances the benefit due to further mixing. Formally, this motivates the assumption due to Flory~\&~Rehner~\cite{flory-jcp-1943a, flory-jcp-1943b} that the nominal free-energy density of the mixture $\mathcal{F}$ is the sum of a stretching contribution and a mixing contribution:
\begin{equation}\label{eq:Fdef}
    \mathcal{F} = \mathcal{F}_\mathrm{stretch} (\lambda_1,\lambda_2,\lambda_3) + \mathcal{F}_\mathrm{mix}(J).
\end{equation}
This is the Helmholtz free energy of the mixture per unit reference volume of dry polymer, where the principal stretches $\lambda_1$, $\lambda_2$, and $\lambda_3$ measure the relative change in linear dimension along the principal axes of the deformation, and the Jacobian determinant $J=\lambda_1\lambda_2\lambda_3$ measures the relative change in bulk volume. We work in terms of principal quantities here for clarity and simplicity; we provide the general version of this theory for arbitrary deformations in Appendices~\ref{app:Eulerian} and~\ref{app:Lagrangian}.

The nominal elastic free-energy density $\mathcal{F}_{\mathrm{stretch}}$ accounts for the stretching of the polymer chains and the nominal free-energy density of mixing $\mathcal{F}_{\mathrm{mix}}$ accounts for the chemical interactions between the polymer chains and the fluid. The former depends on the full deformation field whereas the latter is assumed to be an isotropic function of the local composition, as measured uniquely by $J$ (see Appendices~\ref{app:concentration} and~\ref{app:FloryHuggins}). These two contributions are assumed to be completely independent, which is justified by the very low density of crosslinks, so that the dominant chemical interactions are between the individual monomers and the fluid molecules. These assumptions form the basis for the \textit{theory of ideal elastomeric gels}~\cite{hong-jmps-2008a, cai-epl-2012a, li-softmatter-2012a}.

The mixing contribution $\mathcal{F}_\mathrm{mix}$ measures the free energy of a unit volume of dry polymer after mixing with a volume $J-1$ of fluid, thereby increasing in bulk volume by a factor of $J$. This ignores the elastic penalty of stretching the polymer chains, and would therefore be the same for a polymer solution with no crosslinks. $\mathcal{F}_\mathrm{mix}$ is typically derived from the Flory-Huggins theory of polymer solutions~\cite{flory-jchemphys-1942, huggins-jphyschem-1942, flory-jcp-1943b}, and can be written
\begin{equation}
    \mathcal{F}_\mathrm{mix}(J) =\frac{k_BT}{\Omega_f}\bigg[(J-1)\ln\left(1-\frac{1}{J}\right) -\frac{1}{\alpha}\ln{}J+\chi\left(1-\frac{1}{J}\right)\bigg],
\end{equation}
where $k_B$ is the Boltzmann constant, $T$ is temperature, and $\Omega_f$ is the volume of fluid per fluid molecule in the unmixed state. The first two terms in square brackets reflect the entropy of mixing, where $\alpha$ is a measure of the volume per polymer molecule relative to the volume per fluid molecule in the mixture. The third term reflects the enthalpy of mixing, where $\chi$ is the dimensionless interaction parameter.

The elastic contribution $\mathcal{F}_\mathrm{stretch}$ measures the elastic free energy of a unit volume of dry polymer that has been arbitrarily deformed, ignoring the mixing-related consequences of imbibing or expelling fluid. $\mathcal{F}_\mathrm{stretch}$ is typically derived by assuming a rubber-like (Gaussian-chain) elastic response in the polymer network~\cite{flory-jcp-1943a}, and can be written
\begin{equation}\label{eq:Fstretch}
    \mathcal{F}_\mathrm{stretch}(\lambda_1,\lambda_2,\lambda_3) =\frac{k_BT}{2\Omega_p}\left[\sum_{i=1}^3\lambda_i^2-3-2\ln{}\lambda_1\lambda_2\lambda_3\right],
\end{equation}
where $\Omega_p$ is the volume of polymer per polymer molecule in the unmixed state. This model represents the entropic penalty of deforming a crosslinked network of randomly oriented polymer chains~\citep{boyce-rct-2000a}.

The increase in the total nominal free-energy density of the mixture must be balanced by the external work done. This can be written\footnote{Note that, here and elsewhere, we \textit{do not} adopt the Einstein summation convention.}
\begin{equation}\label{eq:dF}
    \mathrm{d}\mathcal{F} =\sum_{i=1}^3\left(\frac{J\sigma_i}{\lambda_i}\,\mathrm{d}\lambda_i\right) +\frac{\mu_f}{\Omega_f}\mathrm{d}J,
\end{equation}
where $\sigma_i$ are the principal true (Cauchy) total stresses within the mixture and $\mu_f/\Omega_f$ is the chemical potential of the fluid per unit mixture volume, which measures the amount of work required to move a unit volume of fluid from the environment to the mixture. Note that the chemical potential of the polymer does not enter into this balance because the number of polymer chains in the reference volume is fixed by definition. Combining Eq.~\eqref{eq:Fdef} with Eq.~\eqref{eq:dF} and requiring that this remain valid for any arbitrary deformation $\mathrm{d}\lambda_i$, we arrive at a constitutive expression relating $\sigma_i$ to the deformation of the gel,
\begin{equation}\label{eq:totalstress}
    \sigma_i= \frac{\lambda_i}{J}\frac{\partial}{\partial{\lambda_i}}\mathcal{F}_\mathrm{stretch} +\frac{\mathrm{d}}{\mathrm{d}{J}}\mathcal{F}_\mathrm{mix}-\frac{\mu_f}{\Omega_f}.
\end{equation}
We next use these definitions to develop a model for the gel within the framework of poromechanics. 

\subsection{Large-deformation poromechanics}

One classical approach to gel mechanics is based on the theory of linear poroelasticity~\citep{johnson-jchemphys-1982, scherer-jnoncrystsolids-1989, hui-jchemphys-2005}. The resulting ad-hoc models are limited to infinitesimal deformations, and typically neglect the chemical physics of swelling. We generalize this approach by combining the constitutive model for ideal elastomeric gels (discussed above) with the framework of large-deformation poromechanics.

The stretching contribution to the total stress is associated with deformation of the polymer network. In poromechanics, this is known as the Terzaghi effective stress $\sigma^\prime$~\cite{macminn-prapplied-2016},
\begin{equation}\label{eq:sigprime}
    \sigma_i^\prime \equiv\frac{\lambda_i}{J}\frac{\partial}{\partial{\lambda_i}}\mathcal{F}_\mathrm{stretch}=\frac{k_BT}{\Omega_p}\,\left(\frac{\lambda_i^2-1}{J}\right).
\end{equation}
This motivates defining the pore pressure $p$ according to
\begin{equation}\label{eq:pdef}
    p\equiv\frac{\mu_f}{\Omega_f}+\Pi \quad\to\quad \frac{\mu_f}{\Omega_f}=p-\Pi.
\end{equation}
The pore pressure $p$ can then be interpreted as the mechanical contribution to the chemical potential, as usual for an incompressible mixture, and the osmotic pressure $\Pi$ as the mixing contribution,
\begin{equation}\label{eq:PiMix}
    \Pi \equiv-\frac{\mathrm{d}}{\mathrm{d}{J}}\mathcal{F}_\mathrm{mix} =-\frac{k_BT}{\Omega_f}\left[\frac{1}{J}+\ln\left(1-\frac{1}{J}\right)-\frac{1}{\alpha{}J}+\frac{\chi}{J^2}\right].
\end{equation}
Equation~\eqref{eq:totalstress} can then be recast in the familiar form of Biot poroelasticity~\cite{biot-jap-1941a},
\begin{equation}\label{eq:sigdef}
    \sigma_i=\sigma_i^\prime-p.
\end{equation}
The decomposition of total stress into effective stress and pore pressure (Eq.~\ref{eq:sigdef}), and of chemical potential into pore pressure and osmotic pressure (Eq.~\ref{eq:pdef}), is a key distinction between our approach and previous theories~\citep{hong-jmps-2008a, duda-jmps-2010, chester-jmechphyssolids-2010, bouklas-jmps-2015a}. This is central to our interpretation of the mechanics of swelling, allowing us to separate the roles of fluid and solid, and of mechanics and chemistry. Conveniently, this model also reduces to standard poroelasticity when the mixing contribution is negligible~\cite{macminn-prapplied-2016}.

We next outline the main results for spherically symmetric swelling. For clarity, we work strictly in an Eulerian reference frame and in terms of \textit{true} quantities. We provide in Appendices~\ref{app:Eulerian} and~\ref{app:Lagrangian} the general 3D form of the equations, as well as a Lagrangian formulation for comparison.

For a spherically symmetric deformation, the displacement field is purely radial, $\mathbf{u}_s(\mathbf{x},t)=u_s\hat{\mathbf{e}}_r$, and the principal directions are $\hat{\mathbf{e}}_r$, $\hat{\mathbf{e}}_\theta$, and $\hat{\mathbf{e}}_\varphi$. The deformation gradient tensor $\mathbf{F}$ is then diagonal, with principal stretches
\begin{equation}\label{eq:lambdas}
    \lambda_r=\left(1-\frac{\partial{u_s}}{\partial{r}}\right)^{-1} \,\,\mathrm{and}\,\,\, \lambda_\theta=\lambda_\varphi=\left(1-\frac{u_s}{r}\right)^{-1}
\end{equation}
and Jacobian determinant
\begin{equation}\label{eq:J_to_u}
    J=\lambda_r\lambda_\theta\lambda_\varphi=\lambda_r\lambda_\theta^2.
\end{equation}
If the individual densities of the fluid and solid constituents are constant and preserved on mixing, then conservation of volume dictates that $J$ must be related to the local volume fraction of fluid $\phi_f$ (the fluid fraction or porosity) by
\begin{equation}\label{eq:J_to_phi}
    J=\frac{1}{1-\phi_f},
\end{equation}
where we have taken the reference state to be relaxed and dry ($J=1\to\phi_f=0$). Combining Eqs.~\eqref{eq:lambdas}--\eqref{eq:J_to_phi}, we have
\begin{align}\label{eq:u_to_phi}
    \phi_f= \frac{1}{r^2}\frac{\partial}{\partial{r}}\left(r^2u_s-ru_s^2+\frac{1}{3}u_s^3\right),
\end{align}
which can be inverted as
\begin{equation}\label{eq:phi_to_u}
    u_s =r-\left(r^3-3\int_0^r\,r^2\,\phi_f\,\mathrm{d}r\right)^{1/3}.
\end{equation}
Conservation of volume further dictates that
\begin{subequations}\label{eq:phi_to_v}
    \begin{align}
        \frac{\partial{\phi_f}}{\partial{t}} &+\frac{1}{r^2}\frac{\partial}{\partial{r}}\left(r^2\phi_f{}v_f\right)=0\quad{}\mathrm{and} \label{eq:continuity_vf} \\
    \frac{\partial{\phi_s}}{\partial{t}} &+\frac{1}{r^2}\frac{\partial}{\partial{r}}\left(r^2\phi_s{}v_s\right)=0, \label{eq:continuity_vs}
    \end{align}
\end{subequations}
where $\phi_s$ is the local volume fraction of solid, such that $\phi_f+\phi_s=1$, and $v_f$ and $v_s$ are the radial components of the fluid and solid velocities, respectively. Summing Eqs.~\eqref{eq:phi_to_v} and integrating, we have that
\begin{equation}\label{eq:vs_to_vf}
    \phi_f{}v_f+(1-\phi_f)v_s=0,
\end{equation}
which is simply a statement that there is no net flux of material through any cross-section (\textit{i.e.}, in order for fluid to move inward, an equal volume of solid must move outward).

The local flux of fluid relative to the polymer network is driven by gradients in the chemical potential, which accounts for both mechanical and chemical contributions ($p$ and $\Pi$, respectively). This can be written in the form of Darcy's law (see Refs.~\cite{scherer-jnoncrystsolids-1989, peppin-physfluids-2005a} and Appendix~\ref{app:kinetics}),
\begin{equation}\label{eq:darcy}
    \phi_f(v_f-v_s) =-\frac{k(\phi_f)}{\eta}\frac{\partial}{\partial{r}}\left(\frac{\mu_f}{\Omega_f}\right),
\end{equation}
where $k(\phi_f)$ is the deformation-dependent permeability of the solid skeleton, which we take to be an isotropic function of the porosity, and $\eta$ is the dynamic viscosity of the fluid. We adopt a common form for the permeability function~\cite{tokita-jcp-1991a, grattoni-jcis-2001a, engelsberg-pre-2013a},
\begin{equation}\label{eq:permeability}
    k(\phi_f)=k_0\,\frac{\phi_f}{(1-\phi_f)^\beta},
\end{equation}
with characteristic value $k_0$ and parameter $\beta$. References~\cite{tokita-jcp-1991a}, \cite{grattoni-jcis-2001a}, and \cite{engelsberg-pre-2013a} suggest $\beta=1.5$, $1.85$, and $1.75$, respectively. We follow Ref.~\cite{tokita-jcp-1991a}, adopting $\beta=1.5$.

Combining Eqs.~\eqref{eq:phi_to_v}--\eqref{eq:darcy}, we arrive at a conservation law for the porosity in terms of the chemical potential,
\begin{equation}\label{eq:phi_law}
    \frac{\partial{\phi_f}}{\partial{t}} -\frac{1}{r^2}\frac{\partial}{\partial{r}}\left[r^2\,(1-\phi_f)\,\frac{k(\phi_f)}{\eta}\,\frac{\partial}{\partial{r}}\left(\frac{\mu_f}{\Omega_f}\right)\right]=0.
\end{equation}
The chemical potential is then related to the deformation of the solid skeleton by combining Eqs.~\eqref{eq:pdef} and \eqref{eq:sigdef} with mechanical equilibrium, which requires that the divergence of the total stress must vanish. For spherical symmetry, this leads to
\begin{equation}\label{eq:equilibrium}
    \frac{\partial}{\partial{r}}\left(\frac{\mu_f}{\Omega_f}\right) =\frac{\partial{\sigma_r^\prime}}{\partial{r}}+2\frac{\sigma_r^\prime-\sigma_\theta^\prime}{r} -\frac{\partial{\Pi}}{\partial{r}},
\end{equation}
where the radial and azimuthal effective stresses $\sigma_r^\prime$ and $\sigma_\theta^\prime$ ($=$$\sigma_\varphi^\prime$) are provided by taking $i$$=$$r$ and $i$$=$$\theta$, respectively, in Eq.~\eqref{eq:sigprime}. With suitable initial and boundary conditions, we now have an integro-differential system of equations in $\phi_f$, $\mu_f$, and $u_s$ constituting a nonlinear moving-boundary problem.

\begin{figure*}[t!]
    \centering
    \includegraphics[width=17.2cm]{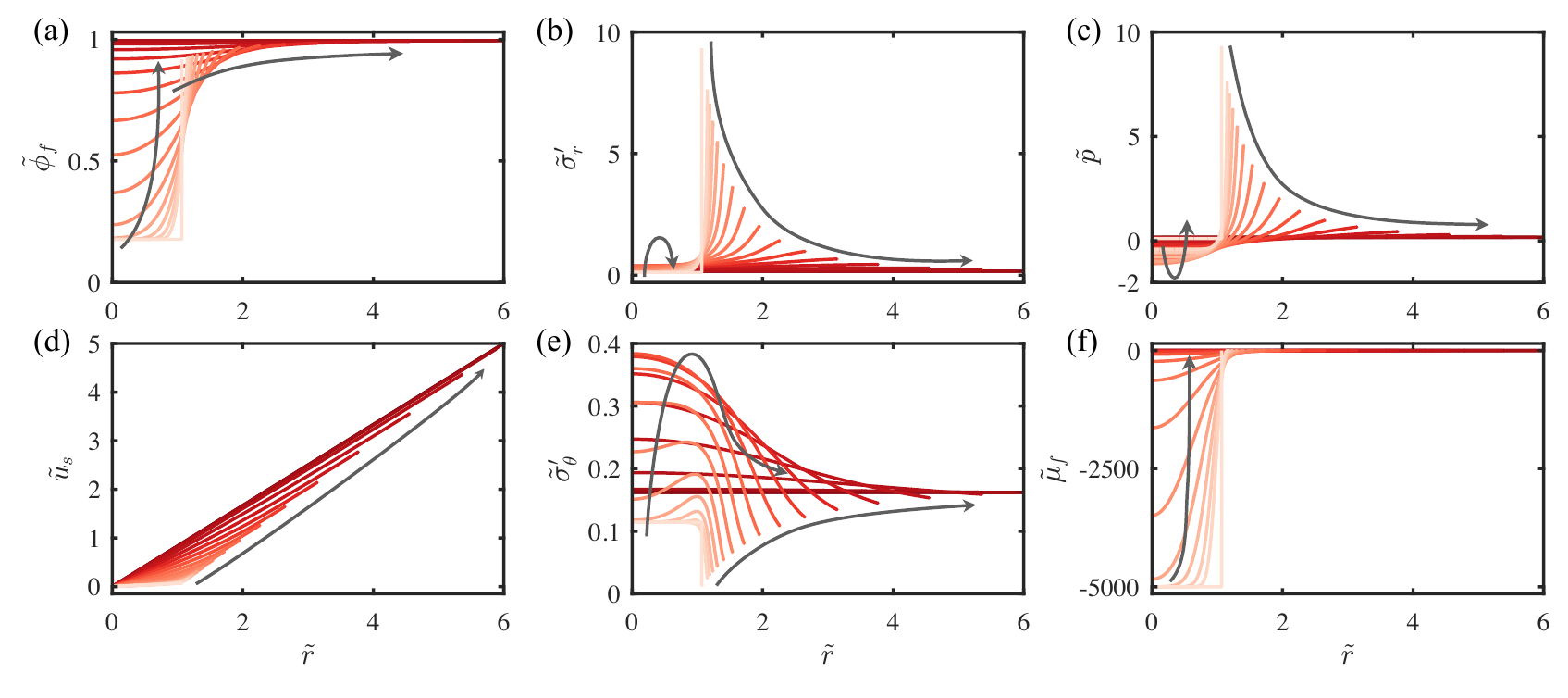}
    \caption{Free swelling: Spatial distributions of (a)~porosity $\phi_f$, (b)~radial effective stress $\tilde{\sigma}_r^\prime$, (c)~pressure $\tilde{p}$, (d)~displacement $\tilde{u}_s$, (e)~azimuthal effective stress $\tilde{\sigma}_\theta^\prime$, and (f)~chemical potential $\tilde{\mu}_f$ (all dimensionless) at $\tilde{t}=0$ and then several times logarithmically spaced between $\tilde{t}=10^{-6}$ and $10^{-1}$ (light to dark~red). The arrows guide the eye through the time evolution, which is in many cases non-monotonic. These results are for material properties $\Omega_f/\Omega_p=1.28\times{}10^{-4}$, $\alpha=250$, and $\chi=0.4$. The initial state is nearly dry ($\tilde{\mu}_{f,0}^{\star}=-5\times{}10^3$ and $\tilde{a}_0=1.067$) and the final state is fully swollen ($\tilde{\mu}_f^\star=0$ and $\tilde{a}_\mathrm{eq}=6$). \label{fig:swelling_model} }
\end{figure*}

\subsection{Scaling}
\label{ss:scaling}

We make the model dimensionless by choosing characteristic time scale $\tau$, length scale $a_d$ (the dry size), permeability scale $k_0$, and stress scale $k_BT/\Omega_p$. We then have, for example,
\begin{align}
    \begin{split}
    \tilde{t}&=\frac{t}{\tau} \,,\quad \tilde{r}=\frac{r}{a_d} \,,\quad \tilde{a}=\frac{a}{a_d} \,,\quad \tilde{u}_s=\frac{u_s}{a_d} \,,\quad \tilde{k}=\frac{k}{k_0} \,, \\
    \tilde{\sigma}_i&=\frac{\sigma_i}{k_BT/\Omega_p} \,,\quad \tilde{\mu}_f=\frac{\mu_f/\Omega_f}{k_BT/\Omega_p} \,,\quad \tilde{\Pi}=\frac{\Pi}{k_BT/\Omega_p}\,,
    \end{split}
\end{align}
where the characteristic time scale is
\begin{equation}\label{eq:tau}
    \tau=\frac{\eta{}a_d^2\Omega_p}{k_0k_BT}.
\end{equation}
The dimensionless model is then fully characterized by just three parameters, which are the three material properties that appear in the dimensionless osmotic pressure,
\begin{equation}\label{eq:PiMix_nondim}
    \tilde{\Pi} =-\frac{\Omega_p}{\Omega_f}\left[\frac{1}{J}+\ln\left(1-\frac{1}{J}\right)-\frac{1}{\alpha{}J}+\frac{\chi}{J^2}\right].
\end{equation}
The dimensionless model is independent of the size of the sphere, implying that swelling is a scale-free process~\cite{matsuo-jcp-1988a}. We continue from this point in dimensionless quantities, which we denote throughout by an over-tilde.

\subsection{Dry state and boundary conditions}

In its fully dry state, the sphere is solid polymer with $\phi_{f,d}=0$. The dry sphere has radius $a_d$ ($\tilde{a}_d=1$) and therefore contains a volume $V_d=\frac{4}{3}\pi{}a_d^3$ of dry polymer. We take the polymer chains to be mechanically relaxed in the dry state, so that
\begin{subequations}
    \begin{align}
        \tilde{u}_{s,d} &= 0, \\
        J_d &= \lambda_{r,d} = \lambda_{\theta,d} = 1, \quad\mathrm{and} \\
        \tilde{\sigma}_{r,d}^\prime &= \tilde{\sigma}_{\theta,d}^\prime = 0.
    \end{align}
\end{subequations}
Relative to this reference state, the sphere will swell to equilibrate its internal chemical potential with that of the surrounding environment. The center of the sphere remains stationary,
\begin{equation}
        \tilde{u}_s(0,\tilde{t}) = \tilde{v}_s(0,\tilde{t}) = \tilde{v}_f(0,\tilde{t}) = 0,
\end{equation}
and the outer boundary of the sphere is a material boundary,
\begin{equation}
    \tilde{u}_s(\tilde{a},\tilde{t}) = \tilde{a}(\tilde{t})-1,
\end{equation}
where $\tilde{a}(t)\geq{}1$ is the radius of the sphere at time $\tilde{t}$. The outer boundary is also unconstrained, so the normal component of the total stress must vanish,
\begin{equation}\label{eq:sigrstar}
    \tilde{\sigma}_r(\tilde{a},\tilde{t})=0 \quad\to\quad \tilde{\sigma}_r^\prime(\tilde{a},\tilde{t}) = \tilde{p}(\tilde{a},\tilde{t}).
\end{equation}
Note that, unlike for a macroscopic porous medium, we cannot impose constraints on $\sigma_r^\prime$ and $p$ individually because the solid and the fluid are mixed at the molecular scale. Lastly, the chemical potential at the outer boundary must always match the ambient value,
\begin{equation}\label{eq:mustar}
    \tilde{\mu}_f(\tilde{a},\tilde{t}) = \tilde{\mu}_f^\star \quad\to\quad \tilde{p}(\tilde{a},\tilde{t})=\tilde{\mu}_f^\star+\tilde{\Pi}(\tilde{a},\tilde{t}),
\end{equation}
where $\tilde{\mu}_f^\star\to-\infty$ gives the fully dry state and $\tilde{\mu}_f^\star=0$ gives the fully swollen state. Note that Eqs.~\eqref{eq:sigrstar} and \eqref{eq:mustar} together imply that the pore pressure is discontinuous across $\tilde{r}=\tilde{a}$, meaning that the pressure just inside the gel always differs from the pressure in the environment.

\subsection{Equilibrium state}

When the sphere reaches equilibrium with its environment, both the fluid and the solid must again be stationary, $\tilde{v}_f=\tilde{v}_s=0$, and the chemical potential must be uniform and equal to the ambient value, $\tilde{\mu}_f=\tilde{\mu}_f^\star$. Equation~\eqref{eq:equilibrium} then provides a nonlinear ordinary differential equation for $\tilde{u}_s$. For an unconstrained sphere (no external stresses), this is satisfied by the isotropic solution
\begin{subequations}
    \begin{align}
        \tilde{u}_s(r) &=[(\tilde{a}_\mathrm{eq}-1)/\tilde{a}_\mathrm{eq}]\,\tilde{r}, \\
        J_\mathrm{eq}=\lambda_r^3 &=\lambda_\theta^3=\tilde{a}_\mathrm{eq}^3, \quad\mathrm{and} \\
        \tilde{\sigma}_r^\prime &=\tilde{\sigma}_\theta^\prime =(\tilde{a}_\mathrm{eq}^2-1)/\tilde{a}_\mathrm{eq}^3.
    \end{align}
\end{subequations}
The equilibrium radius $\tilde{a}_\mathrm{eq}$ is determined by the nonlinear algebraic equation $\tilde{\sigma}_r^{\prime}(\tilde{a}_\mathrm{eq})=\tilde{\Pi}(\tilde{a}_\mathrm{eq}^3)+\tilde{\mu}_f^\star$. The result depends only on $\tilde{\mu}_f^\star$ and the three dimensionless material properties: $\Omega_f/\Omega_p$, $\alpha$, and $\chi$ (Eq.~\ref{eq:PiMix_nondim}).

\section{Dynamics of swelling}
\label{s:swelling}

A hydrogel sphere that is initially at equilibrium with ambient chemical potential $\tilde{\mu}^\star_{f,0}$ will swell when exposed to a new chemical potential $\tilde{\mu}^\star_f>\tilde{\mu}^\star_{f,0}$. Swelling will stop when the sphere reaches equilibrium with its new environment.

\subsection{Poromechanics of swelling}\label{ss:swelling_model}

We consider a sphere that is initially at equilibrium with air of relative humidity $\mathrm{RH}\approx{}0.6$, for which the sphere is nearly dry. The chemical potential in this initial state is then $\tilde{\mu}_f(\tilde{r},0)=\tilde{\mu}^\star_{f,0}=(\Omega_p/\Omega_f)\ln(\mathrm{RH})$. At $\tilde{t}=0^+$, the sphere is suddenly immersed in water, for which $\tilde{\mu}_f^\star\approx{}0\gg{}\tilde{\mu}_{f,0}^\star$. The final state will be a new equilibrium state at which $\tilde{\mu}_f(\tilde{r},\tilde{t})\to{}\tilde{\mu}_f^\star$. We study the dynamics of this transition numerically using a finite-volume method with an adaptive grid and explicit time integration~(see Appendix~\ref{app:Numerics}). Typical results are shown in Fig.~\ref{fig:swelling_model}.

The displacement $\tilde{u}_s$ is strictly positive, meaning that all material points move strictly radially outward from their initial positions throughout the swelling process~(Fig.~\ref{fig:swelling_model}b). However, there is also a positive and increasing gradient in displacement from the center to the outer edge, indicating that material points near the outer radius move outward earlier and further than those closer to the center. This is indicative of strongly nonuniform volumetric expansion in a spherical geometry. Accordingly, we find that the porosity $\phi_f$ near the outer boundary increases sharply at early times as the dry gel on the outside rapidly imbibes water (Fig.~\ref{fig:swelling_model}a). This rapid swelling of the outer region is inhibited by its attachment to the comparatively unswollen core, leading to a strongly tensile radial effective stress in the outer region that relaxes as the swelling process proceeds inward~(Fig.~\ref{fig:swelling_model}b). 

The pore pressure just inside the gel must exceed the ambient pressure by the osmotic pressure throughout the swelling process, and at equilibrium ($\tilde{p}(\tilde{a},\tilde{t})=\tilde{\Pi}(\tilde{a},\tilde{t})$ from  Eq.~\ref{eq:mustar}). Fluid flows into the gel from the environment despite this larger-than-ambient pore pressure because flow is in the direction of decreasing chemical potential $\tilde{\mu}_f$, and the chemical potential decreases monotonically toward the center. This gradient becomes gentler as the chemical potential throughout as the gel increases, equilibrating with the ambient value. 

The effective stresses everywhere are strictly positive (tensile) throughout the swelling process since the polymer chains are being stretched to accommodate additional pore fluid~(Fig.~\ref{fig:swelling_model}b,e). The mechanical support for this stretching is provided by the large pore pressure~(Fig.~\ref{fig:swelling_model}c). The gel behaves in this sense like an inflating balloon, with pressure in the fluid balancing elastic stretching in the solid, the distinction being that this is a bulk phenomenon within the gel.

Although the azimuthal effective stress $\tilde{\sigma}_\theta^\prime$ is tensile everywhere, the azimuthal total stress $\tilde{\sigma}_\theta$ is strongly compressive in the outer region where the pore pressure far exceeds the tensile effective stress. This reflects the fact that the outer region is imbibing fluid and trying to grow while being bonded to the unswollen inner region.

\subsection{Swelling experiments}\label{ss:swelling_experiments}

\begin{figure*}[t!]
    \centering
    \includegraphics[width=17.2cm]{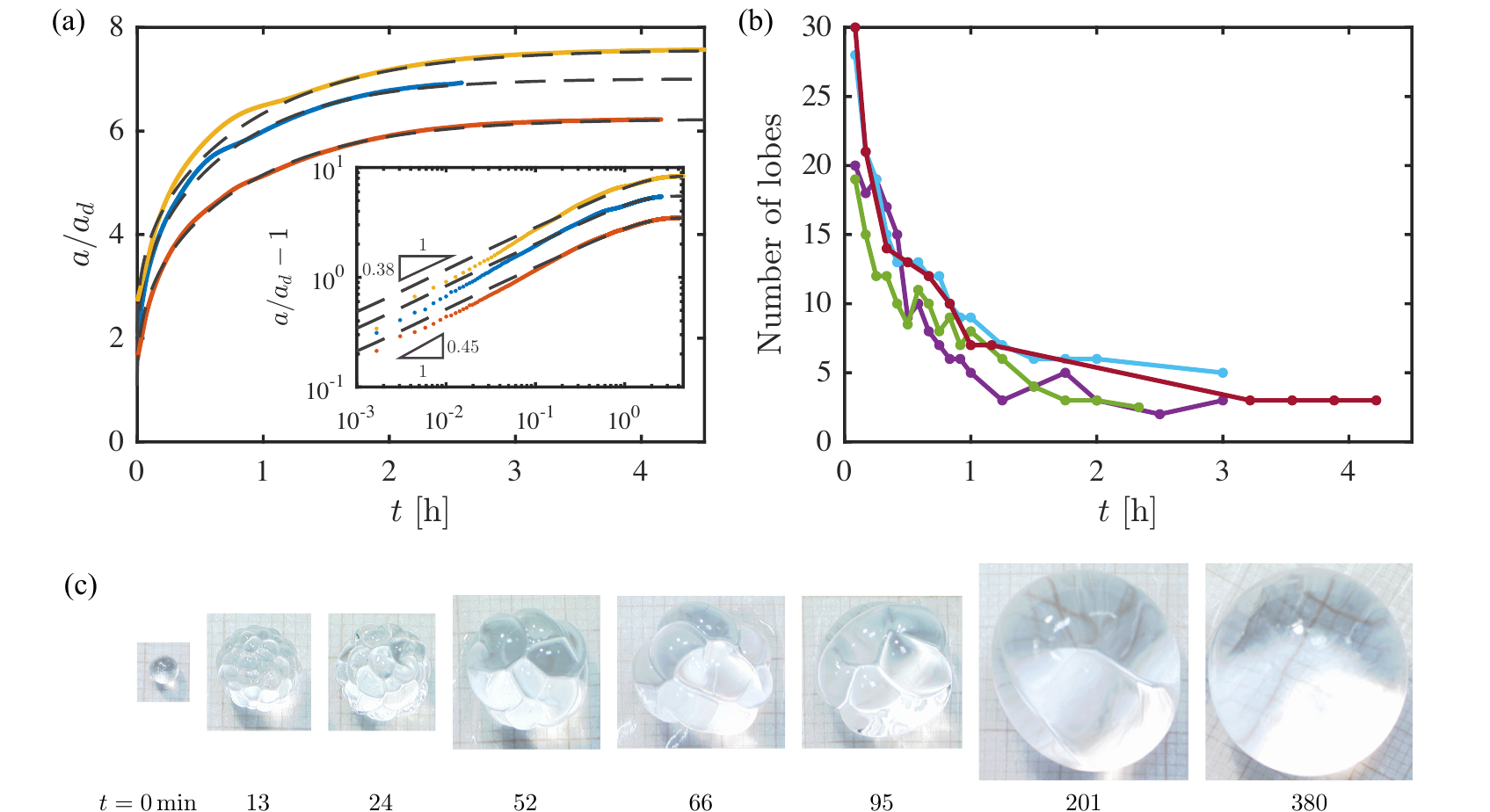}
    \caption{The swelling of a spherical gel. (a)~Time evolution of the radii of three hydrogel spheres after immersion in water, showing experimental data (orange, blue, and yellow, shifted vertically by 0, 0.5, and 1, respectively, for clarity) and the predictions of the model (dashed gray, also shifted by the same amounts). The inset shows $a/a_d-1$ against $t$ for the data and the model on a logarithmic scale to highlight the power-law behavior at early times (same colors, and scaled vertically by factors of 2/3, 1, and 3/2, respectively, for clarity). (b)~Time evolution of the number of lobes around a circumference of the swelling sphere for four different experiments. (c)~Photographs of a swelling gel at different times, as indicated, where the initial radius is $a_d\approx{}1.5\,\mathrm{mm}$ and the final radius is $\sim$$6.7a_d$. \label{fig:swelling_experiment} }
\end{figure*}

To study swelling experimentally, we submerge dry polyacrylamide hydrogel beads (Educational Innovations) in a container of water (Volvic or EMD~Millipore) and photograph them at regular time intervals using a digital camera. Via image processing, we then extract the average radius of the bead and the number of lobes around the circumference, both in the plane of the image (Fig.~\ref{fig:swelling_experiment}).

We show the time evolution of the average radius, $a/a_d$, in Fig.~\ref{fig:swelling_experiment}a for three different beads. To compare these results with the model, we need to determine the three material properties $\alpha$, $\chi$, and $\Omega_f/\Omega_p$, as well as the dry size $a_d$ for each bead. The material properties are unknown and difficult to measure directly. For all three beads, we adopt $\alpha=250$ and $\chi=0.4$, similar to values used for similar materials in previous studies~\cite{engelsberg-pre-2013a}. We further assume $\mathrm{RH}=0.6$ in the initial state. We can then calculate the dry sizes of the beads, which are essentially independent of $\Omega_f/\Omega_p$~(see Appendix~\ref{app:drysize}). Finally, we use $\Omega_f/\Omega_p$ as a fitting parameter to match the final equilibrium size of each bead, which leads to $\Omega_f/\Omega_p\sim{}1.09\times{}10^{-4}$ with a variation between beads of roughly $\pm{}7\%$. Note that variation in material properties has been noted previously, even within the same batch~\cite{matsuo-jcp-1988a}. The dimensionless swelling problem is then fully specified.

To plot the model results against dimensional time, we need to calculate the characteristic time scale $\tau$ (Eq.~\ref{eq:tau}). To do so, we take $\Omega_f=2.99\times{}10^{-29}\,\mathrm{m}^{3}$, $\eta=10^{-3}\,\mathrm{Pa}\,\mathrm{s}$, $k_B=1.38\times{}10^{-23}\,\mathrm{J}\,\mathrm{K}^{-1}$, and $T=295\,\mathrm{K}$. The final quantity in the time scale is the characteristic permeability $k_0$; we choose the value for which the model best matches the experiment, $k_0=8.0\times{}10^{-20}\,\mathrm{m}^2$. This value is again similar to that used in previous work~\cite{tokita-jcp-1991a, engelsberg-pre-2013a}. We use this value for all beads. The associated characteristic times are $\tau\sim{}4.5\times{}10^{5}\,\mathrm{s}$, with a variation of about 20\% due to the slightly different dry sizes. Having fitted the final radius and calculated the time scale, the model provides a good qualitative and quantitative match with the data. The number of unknown parameters is sufficiently large that no particular set of values can be said to provide a unique match, but these values provide a useful comparison.

The inset of Fig.~\ref{fig:swelling_experiment}a highlights the early-time evolution, indicating a power-law growth of the form $(a/a_d-1)\propto{}t^{0.45}$, suggesting that swelling is dominated by diffusion-like transport of water into the gel at very early times. The model also follows a power law at early times, but with an exponent closer to $0.38$. The discrepancy may be due to the surface instability, which leads to a large change in the surface area of the bead and may fundamentally change the dynamics of swelling.

Other than the extreme increase in volume, the most striking aspect of swelling is the development and evolution of the lobe-like surface pattern, a well-known phenomenon~\cite{tanaka-n-1987a, dervaux-prl-2011a, barros-softmatter-2012a}. 

\begin{figure*}
    \centering
    \includegraphics[width=17.2cm]{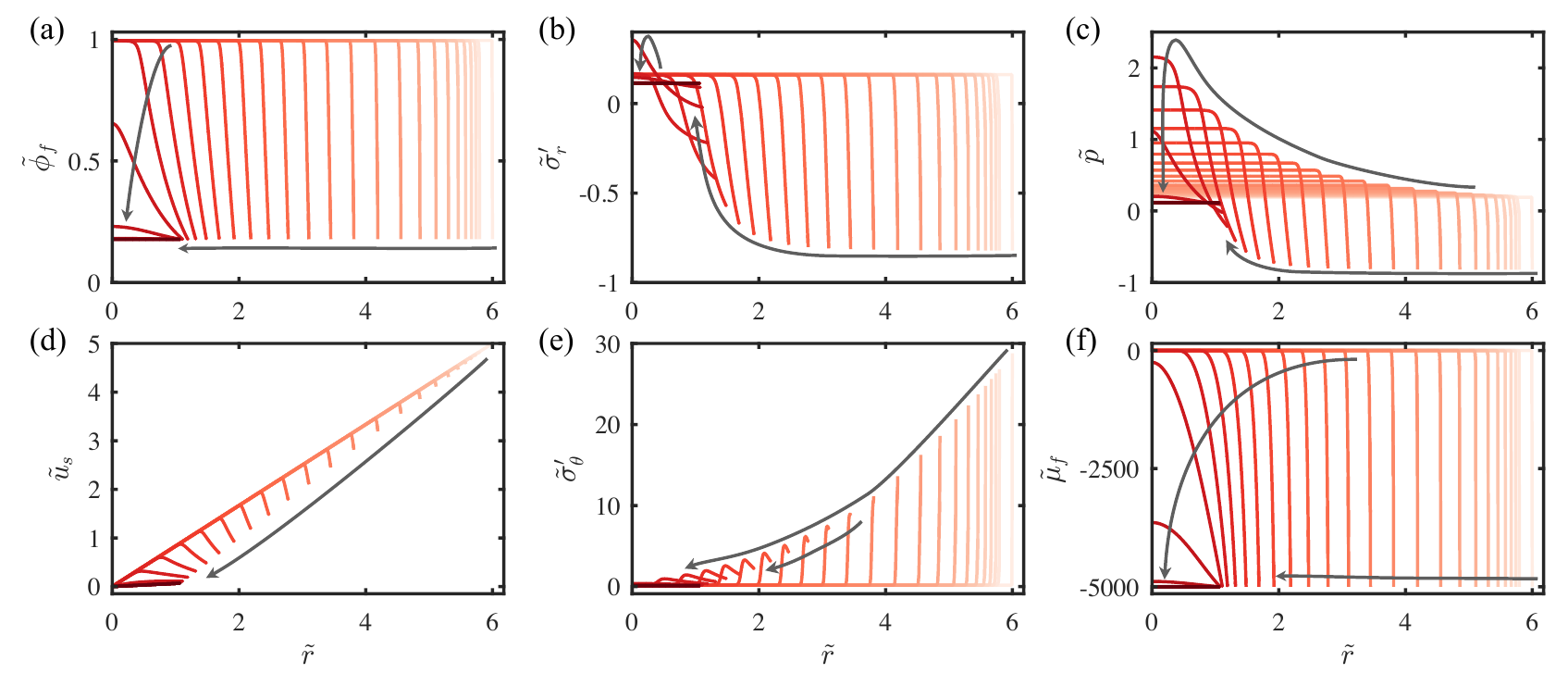}
    \caption{Free drying: Spatial distributions of (a)~porosity $\phi_f$, (b)~radial effective stress $\tilde{\sigma}_r^\prime$, (c)~pressure $\tilde{p}$, (d)~displacement $\tilde{u}_s$, (e)~azimuthal effective stress $\tilde{\sigma}_\theta^\prime$, and (f)~chemical potential $\tilde{\mu}_f$ (all dimensionless) at $\tilde{t}=0$ and then several times logarithmically spaced between $\tilde{t}=10^{-12}$ and $10^{-3}$ (light to dark~red). These results are for the same material properties used in Fig.~\ref{fig:swelling_model}, but the ambient conditions and the initial and final states are precisely reversed. The arrows guide the eye through the time evolution, which is strikingly different from swelling (\textit{cf.}~Fig.~\ref{fig:swelling_model}). \label{fig:drying_model} }
\end{figure*}

\subsection{Transient surface instability}\label{ss:instability}

Interfacial growth has long been linked to pattern formation~\cite{eden-proceedings-1961a, kardar-prl-1986a, family-physicaa-1990a}. More recently, volumetric growth under fixed, external constraints has attracted attention due to its likely role in developmental morphogenesis~\cite{ben-amar-jmps-2005a, yin-pnas-2008a, ciarletta-prl-2014a, tallinen-pnas-2014a}. In swelling, the fluid content provides an evolving internal constraint that can lead to the formation of both steady and transient patterns~\cite{tanaka-n-1987a, dervaux-prl-2011a, barros-softmatter-2012a, tallinen-natphys-2016}.

For a bead of initial radius $\sim$$1.5\,\mathrm{mm}$, the swelling process takes about $5\,\mathrm{h}$ (Fig.~\ref{fig:swelling_experiment}c). During this time, the surface of the bead exhibits a transient pattern that evolves from small-scale to large-scale features through a coarsening process where neighboring lobes grow and merge (Fig.~\ref{fig:swelling_experiment}b). Small-scale surface roughness emerges and then rapidly develops into a relatively uniform tiling of hexagonal lobes~(Fig.~\ref{fig:swelling_experiment}c, $13$--$52\,\mathrm{min}$). This pattern transitions to a randomly oriented network of folds or wrinkles at later times (Fig.~\ref{fig:swelling_experiment}c, $52$--$201\,\mathrm{min}$), and these ultimately merge and fade back into a smooth spherical surface (Fig.~\ref{fig:swelling_experiment}c, $380\,\mathrm{min}$). The bead then continues to grow smoothly until reaching its equilibrium size.

As described in the previous section, swelling is characterized by a rapidly growing outer shell that is constrained by a relatively unswollen inner core~(see Appendix~\ref{app:coreshell}). The shell is soft relative to the comparatively unswollen core, and the surface pattern has been attributed to the development of compressive azimuthal stress in the shell due to its attachment to the core~\cite{tanaka-n-1987a}. We have provided quantitative evidence for this compressive stress (Fig.~\ref{fig:swelling_model}b,e and Appendix~\ref{app:stressmap}), which is ultimately a result of the strongly heterogeneous fluid content in the bead at early times. The fact that the lobes result from a mechanical constraint implies that they would disappear if the constraint were removed; indeed, we find that the lobes disappear locally when a lobed bead is sliced with a blade. The fact that the lobes result from heterogeneous water content further implies that the lobes would gradually vanish if a partially swollen bead were removed from water, allowing the water content to equilibrate within the bead; we have verified this experimentally.

The wavelength of the lobes is roughly proportional to the thickness of the soft shell, which is the relevant length scale for the instability~\cite{barros-softmatter-2012a}. However, this is not as simple as a compressed soft layer bonded to a rigid substrate~\cite{yin-pnas-2008a, li-prl-2011a, ciarletta-prl-2013a, tallinen-pnas-2014a}; it is a single material with a continuous stiffness distribution, where the thicknesses and stiffnesses of both layers, as well as the compressive total stress that drives the instability, all evolve with time. This suggests that the instability cannot be understood in isolation from the dynamics of swelling.

\section{Dynamics of drying}
\label{s:drying}

In hydrogels, swelling is reversible. However, the reverse process---de-swelling or drying---has received little attention. We now consider the fate of a fully swollen hydrogel bead that is suddenly removed into air. The bead will subsequently shrink until it reaches equilibrium with its new environment.

\subsection{Poromechanics of drying}
\label{ss:drying_model}

To illustrate the physics of drying, we consider the reversal of the swelling process shown in Fig.~\ref{fig:swelling_model} for an identical sphere (same size and material properties). The sphere is initially fully swollen ($\tilde{a}_0=6$ for $\tilde{\mu}_{f,0}^\star=0$) and, at $\tilde{t}=0^+$, it is suddenly removed to a dry environment with corresponding ambient chemical potential $\tilde{\mu}_f^\star\ll{}\tilde{\mu}_{f,0}^\star$, which provides the incentive for drying. The final state will be a new equilibrium state in which the sphere is nearly dry ($\tilde{a}_\mathrm{eq}=1.07$ for $\tilde{\mu}_f^\star=-5\times{}10^3$). We solve the problem numerically, as before, and typical results are shown in Fig.~\ref{fig:drying_model}.

We find that the transient evolution during drying is strikingly different from swelling, despite the fact that the ambient conditions and the initial and final states are precisely reversed from swelling. This is a signature of the nonlinearity of large deformations---for small deformations, drying is essentially a mirror image of swelling (see Appendix~\ref{app:smallswelling}).

Drying propagates inward over time as a sharp drying front. Behind (outward of) this front is a thin outer region in which the polymer chains are in strong azimuthal tension (Fig.~\ref{fig:drying_model}e). Ahead of (inward of) this front is a quiescent core in which everything except the pressure remains at its initial value until the front arrives. The pressure ahead of the front rises uniformly and monotonically as the front progresses inward (Fig.~\ref{fig:drying_model}c). This reflects the fact that the fluid within the gel is being squeezed by the tight and contracting outer shell---the elevated pressure is the mechanical response to this squeezing, providing the outward force that supports the tensile azimuthal stress in the shell. Once the drying front arrives at the center, all quantities decay smoothly toward their final values.

\subsection{Drying experiments}
\label{ss:drying_experiments}

\begin{figure*}[t!]
    \centering
    \includegraphics[width=17.2cm]{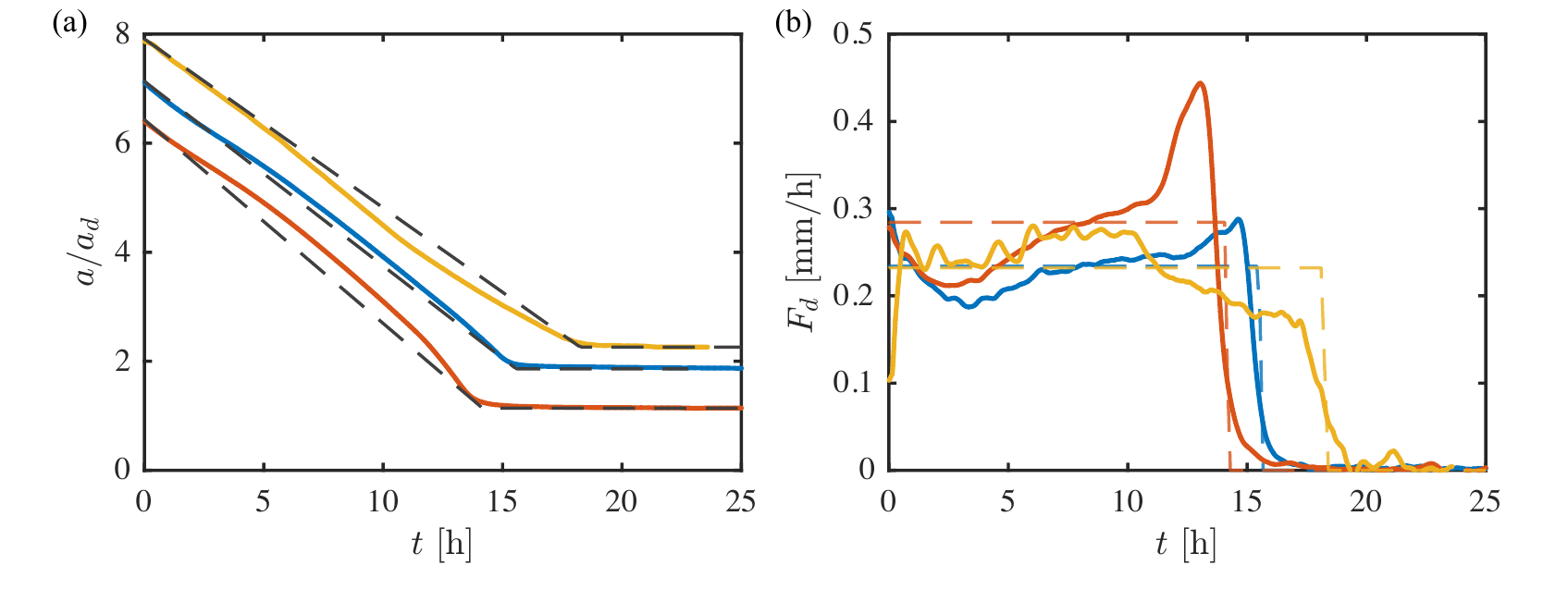}
    \caption{The drying of a spherical gel. (a)~Time evolution of the radii of three different gel beads after removal from water to air, showing experimental data (blue, orange, and yellow, shifted vertically by 0, 0.66, and 1, respectively, for clarity) and the prediction of the model (dashed gray). (b)~Time evolution of the drying flux from the experiments and the model. The experiments exhibit a drying flux that is nearly constant in time, $F_d^\star\approx{}$0.284, 0.234, and 0.232~mm$\,$h$^{-1}$ ($\tilde{F}_d^\star\approx{}$50.8, 38.6, and 48.2, respectively), with some small variations that may be due to variation in the ambient relative humidity over the long duration of drying, or because the assumption of a constant drying rate is a crude approximation to the true dynamics of water transport in the room. Swelling data for these beads is shown in Fig.~\ref{fig:swelling_experiment} (same colors) and we determine the best-fit material properties from the swelling results. \label{fig:drying_experiment} }
\end{figure*}

To study drying experimentally, we remove fully swollen hydrogel beads into air and photograph them at regular time intervals using a digital camera. Macroscopically, the most striking aspect of drying is the lack of a surface instability---the gel remains smooth and spherical throughout the drying process. This observation is supported by the model, which shows azimuthal tension rather than compression in the outer layer of the gel. Other authors have observed patterns during de-swelling in gels experiencing a sharp chemically or thermally induced phase transition~(\textit{e.g.}, \cite{matsuo-jcp-1988a, matsuo-nature-1992, boudaoud-pre-2003}). In our system, swelling and de-swelling are driven by sudden changes in the ambient chemical potential, which leads to a smooth evolution of the gel structure, and it is not entirely surprising that this leads to qualitatively different behavior. We do not consider thermally induced swelling here since our experiments are approximately isothermal, but it can be readily introduced by adopting $\chi=\chi(T)$~\cite{doi-jpsjpn-2009}.

We plot the time evolution of the average radius for three beads in Fig.~\ref{fig:drying_experiment}a, and we find that this decreases roughly linearly with time in all cases. To explain this observation, we consider the evolution of the drying flux $F_d$, which is the flux of water exiting the bead at the surface. Conservation of volume dictates that this must be given by
\begin{equation}
    F_d\equiv{}-\frac{1}{4\pi{}a^2}\,\frac{\mathrm{d}}{\mathrm{d}t} \left(\frac{4}{3}\pi{}a^3\right) =-\frac{\mathrm{d}a}{\mathrm{d}t} \quad\to\quad \tilde{F}_d=-\frac{\mathrm{d}\tilde{a}}{\mathrm{d}\tilde{t}}
\end{equation}
We can therefore calculate $F_d$ directly from the experimental measurements and from the model (Fig.~\ref{fig:drying_experiment}b). In the absence of other constraints, the drying flux evolves naturally with the rate of internal water transport to the surface of the bead. We refer to drying under these conditions as ``free drying''.

We find that free drying is much faster than swelling. Swelling is resisted by the elastic stress in the polymer chains, which must be stretched to expand the pore space; drying, in contrast is accelerated by the relaxation of elastic stress in the polymer chains, which helps to squeeze water out of the bead. For the beads shown Fig.~\ref{fig:drying_experiment}, the model predicts that these beads would dry completely in a matter of minutes under free-drying conditions~(see Appendix~\ref{app:freedrying}), but our experiments take $\sim$$15\,\mathrm{h}$. This demonstrates clearly that the experiments are not in a state of free drying.

The drying flux in the experiments can also be constrained externally by the rate of water transport away from the surface of the bead since residual water will shield the bead from the true ambient chemical potential. In our experiments, this water transport occurs by evaporation. The linear decrease of the radius with time suggests that the drying flux due to evaporation is roughly constant. To account for this constraint in the model, we assume that ambient conditions lead to a maximum evaporation rate $F_d^\star$. When the natural drying rate $F_d(t)$ would otherwise exceed $F_d^\star$, we assume that excess moisture accumulates on the outside of the bead or in the air, shielding the bead from the true ambient chemical potential $\mu_f^\star$. We impose this as a constraint by dynamically adjusting $\mu_f^\star$ to ensure that $F_d(t)\leq{}F_d^\star$. Measuring $F_d^\star$ from our experiments, we find that this model is indeed able to reproduce the dynamics of evaporation-limited drying (Fig.~\ref{fig:drying_experiment}a,b).

We use the model to study evaporation-limited drying in more detail, presenting results for several values of $F_d^\star$ in Fig.~\ref{fig:drying_model_Fd} (see also, Appendix~\ref{app:drying_constrained}). For finite $F_d^\star$, drying of a swollen bead takes place in two stages. At early times, the radius of the bead decays linearly with time (Fig.~\ref{fig:drying_model_Fd}a). The slope of this linear regime is controlled by $F_d^\star$, as evidenced by the plateau in the flux at early times~(Fig.~\ref{fig:drying_model_Fd}b). We show in the inset of Fig.~\ref{fig:drying_model_Fd}b the values of the flux at $t=0$ as a function of $F_d^\star$. At later times, the radius decreases more slowly and eventually saturates to an equilibrium state (Fig.~\ref{fig:drying_model_Fd}a). The crossover times for the various values of $F_d^\star$ are marked on Fig.~\ref{fig:drying_model_Fd} as vertical dashed lines. Physically, this transition can be understood as a crossover between an early regime where drying is limited by water transport away from the bead, so that the drying dynamics are controlled by the ambient conditions through the value of $F_d^\star$, to a late regime where drying is limited by water transport within the bead. As the bead dries, the porosity field becomes increasingly heterogeneous~(Fig.~\ref{fig:drying_model}c). In particular, its outermost layer shows a very low porosity compared to its center. As the porosity decreases, so does the typical pore size. Thus, it becomes increasingly hard for water molecules to reach the surface. We find evidence of this in the agreement between the crossover time scale measured from Fig.~\ref{fig:drying_model_Fd}b and the time at which the porosity reaches its equilibrium value at the surface of the bead, as shown on Fig.~\ref{fig:drying_model_Fd}c.

\subsection{Fracture during drying}
\label{ss:fracture}

\begin{figure*}[t!]
    \centering
    \includegraphics[width=17.2cm]{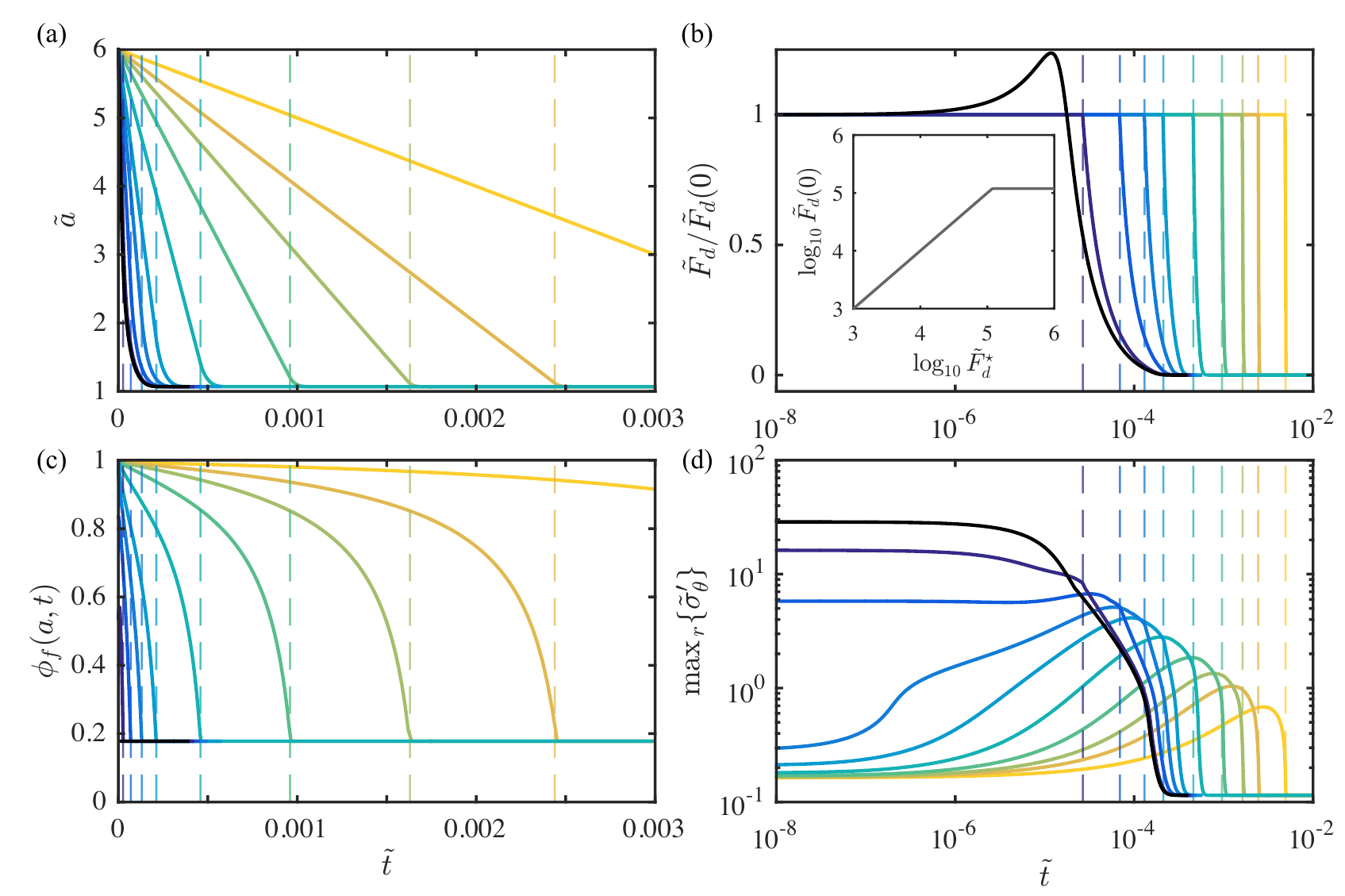}
    \caption{Using the model, we study evaporation-limited drying: (a)~Evolution of the outer radius $\tilde{a}$, (b)~drying flux $\tilde{F}_d$, (c)~porosity at the outer radius $\phi_f(\tilde{a},\tilde{t})$, and (d)~maximum azimuthal stress $\max_{\,r}\{\tilde{\sigma}_\theta^\prime\}$ for $\tilde{F}_d^\star\to\infty$ (free drying, black line) and then for nine values logarithmically spaced between $\tilde{F}_d^\star=1\times{}10^5$ and $1\times{}10^3$ (dark to light colors). Note that free drying exhibits a maximum drying rate of about $1.5\times{}10^5$ for these parameters, so any value of $\tilde{F}_d^\star$ greater than this would be equivalent to free drying. \label{fig:drying_model_Fd} }
\end{figure*}

Evaporation-limited drying involves a competition between water transport within the bead and water transport away from the bead. In free drying and for large $\tilde{F}_d^\star$, water initially escapes the surface of the bead much faster than it can diffuse through the pore structure and the water content becomes highly heterogeneous. This leads to large internal tensile stresses with a maximum value close to the surface, and this maximum stress increases with $\tilde{F}_d^\star$. For strongly limited drying (small $\tilde{F}_d^\star$), the water content within the bead is less heterogeneous because the water has more time to redistribute. At very low values of $\tilde{F}_d^\star$, the water content within the bead is nearly homogeneous and drying can be captured with a quasi-static model (see Appendix~\ref{app:QSmodel}). We plot the time evolution of the maximum azimuthal stress within the bead $\max_{\,r}\{\tilde{\sigma}^\prime_\theta\}$ for various values of $\tilde{F}_d^\star$ in Fig.~\ref{fig:drying_model_Fd}d. This maximum occurs at $t=0$ for large $\tilde{F}_d^\star$, but decreases and then shifts to later times as $\tilde{F}_d^\star$ decreases.

\begin{figure}
    \centering
    \includegraphics[width=8.6cm]{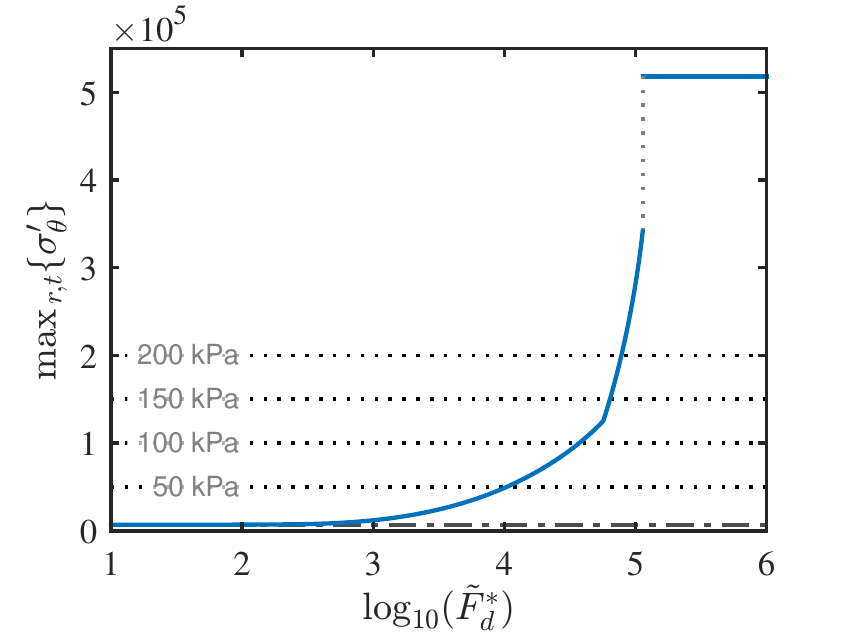}
    \caption{ Fracture during drying. Dimensional overall maximum azimuthal stress experienced by the bead during drying $\max_{\,r,t}\{\sigma^\prime_\theta\}$ as a function of the dimensionless maximum evaporation rate $\tilde{F}_d^\star$. The black dash-dotted line represents the overall maximum azimuthal stress level in quasi-static drying for $\tilde{F}_d^\star\ll{}10^2$. For $\tilde{F}_d^\star \geq 1.2\times{}10^5$, $\max_{\,r,t}\{\sigma^\prime_\theta\}$ jumps to its free-drying value. We plot as horizontal black dotted lines typical values of fracture stresses for polyacrylamide hydrogels~\cite{gong-am-2003a}.
    \label{fig:drying_model_fracture} }
\end{figure}

We plot the overall maximum azimuthal stress during drying $\max_{\,r,t}\{\tilde{\sigma}^\prime_\theta\}$ as a function of $\tilde{F}_d^\star$ in Fig.~\ref{fig:drying_model_fracture}. The overall maximum stress increases with $\tilde{F}_d^\star$ from a minimum value in the quasi-static limit ($\max_{\,r,t}\{\tilde{\sigma}^\prime_\theta\}=0.385$ for $\tilde{F}_d^\star\ll{}10^2$) to a maximum value in the free-drying limit ($\max_{\,r,t}\{\tilde{\sigma}^\prime_\theta\}=29.8$ for $\tilde{F}_d^\star>1.2\times{}10^5$). The curve has a noticeable discontinuity in its slope near $\tilde{F}_d^\star=5\times{}10^4$, to the right of which the overall maximum stress occurs at $t=0$ and to the left of which this occurs at later times. For free drying, the initial evaporation rate is $\tilde{F}_d(0) \approx 1.2\times{}10^5$ (Fig.~\ref{fig:drying_model_Fd}b), and this then grows to a maximum value of $\tilde{F}_d \approx 1.5\times{}10^5$ before declining monotonically to zero. For the range $1.2\times{}10^5<\tilde{F}_d^\star<1.5\times{}10^5$, $\tilde{F}_d(0)$ is then insensitive to $\tilde{F}_d^\star$ since drying is not limited by evaporation until $\tilde{F}_d(t)$ reaches $\tilde{F}_d^\star$. As a result, the overall maximum stress jumps to its free-drying value near $\tilde{F}_d^\star=1.2\times{}10^5$, which is in the range where the overall maximum stress occurs at $t=0$ and the initial drying behavior is not limited by evaporation. Drying is completely free for $\tilde{F}_d^\star>1.5\times{}10^5$. As a consequence, a plateau develops in the overall maximum stress for $\tilde{F}_d^\star>1.2\times{}10^5$, and this plateau takes the value corresponding to free drying. The resulting range of stresses spans two orders of magnitude and can readily exceed the typical fracture stress of hydrogels~(Fig.~\ref{fig:drying_model_fracture}). Although our drying experiments are well below the fracture threshold (\textit{cf.}, Figs.~\ref{fig:drying_experiment} and \ref{fig:drying_model_fracture}), we have verified experimentally that accelerated drying can indeed result in fracture. A detailed experimental investigation of drying-induced fracture is beyond the scope of the present study, but will be the subject of future work. Fracturing due to the development of heterogeneous water content is also well known as a pattern-forming process in drying suspensions~\cite{dufresne-langmuir-2006a, goehring-pnas-2009a}.

\section{Conclusions}
\label{s:conclusions}

Hydrogels are remarkable porous materials that can exhibit extreme but reversible changes in volume by imbibing or expelling hundreds of times their own weight in water in response to external stimuli. Hydrogels have great potential in applications ranging from sensing to drug delivery, and are already widely used in applications such as moisture absorption and soft contact lenses. A clear understanding of the dynamics of swelling and drying is essential for engineering design, from optimising the rate of drug release to avoiding cracking in reusable sensors or actuators, but the vast majority of previous work on gels has focused on their equilibrium chemical physics, or has been limited to relatively small volume changes.

Beginning with the theory of ideal elastomeric gels, we have provided a concrete poromechanical interpretation for swelling and drying by introducing the classical Terzaghi decomposition of total stress into effective stress and pore pressure. We have provided a detailed exploration of the internal mechanics of these processes, as well as a quantitative comparison between experiment and theory for the dynamics of swelling and drying, with the gel increasing or decreasing in volume by a factor of about 200. In doing so, we have highlighted the striking and transient differences between swelling and drying. An important implication of our results is that both the compressive total stresses during swelling and the tensile effective stresses during drying can be minimized by swelling or drying slowly, as demonstrated by our quantitative investigation of the role of external constraints on the drying rate and their implications for fracturing during drying.

This study is an important step toward understanding the transient mechanics of swelling and drying. In particular, a clear direction for future work is the exploration of swelling and drying in 3D, which would allow for other geometries and for capturing the elastic instability. We highlighted the role of the evaporation rate on the risk of fracture during drying, but much is left to explore in terms of the other parameters of the model. For example, the impact of different solvents and the presence of other solutes are central to applications in biomedical engineering. The framework described here will also be useful for understanding swelling driven by other environmental stimuli, such as temperature, with relevance to biological processes and industrial applications.



TB was supported in part by the Yale School of Engineering \& Applied Science Advanced Graduate Leadership Program. JP acknowledges the assistance of Natacha Mac\'{e} and Norio Yonezawa for assistance in some of the experiments. SM was supported in part by NSF-DMR~1410157.


\begin{thebibliography}{58}%
\makeatletter
\providecommand \@ifxundefined [1]{%
 \@ifx{#1\undefined}
}%
\providecommand \@ifnum [1]{%
 \ifnum #1\expandafter \@firstoftwo
 \else \expandafter \@secondoftwo
 \fi
}%
\providecommand \@ifx [1]{%
 \ifx #1\expandafter \@firstoftwo
 \else \expandafter \@secondoftwo
 \fi
}%
\providecommand \natexlab [1]{#1}%
\providecommand \enquote  [1]{``#1''}%
\providecommand \bibnamefont  [1]{#1}%
\providecommand \bibfnamefont [1]{#1}%
\providecommand \citenamefont [1]{#1}%
\providecommand \href@noop [0]{\@secondoftwo}%
\providecommand \href [0]{\begingroup \@sanitize@url \@href}%
\providecommand \@href[1]{\@@startlink{#1}\@@href}%
\providecommand \@@href[1]{\endgroup#1\@@endlink}%
\providecommand \@sanitize@url [0]{\catcode `\\12\catcode `\$12\catcode
  `\&12\catcode `\#12\catcode `\^12\catcode `\_12\catcode `\%12\relax}%
\providecommand \@@startlink[1]{}%
\providecommand \@@endlink[0]{}%
\providecommand \url  [0]{\begingroup\@sanitize@url \@url }%
\providecommand \@url [1]{\endgroup\@href {#1}{\urlprefix }}%
\providecommand \urlprefix  [0]{URL }%
\providecommand \Eprint [0]{\href }%
\providecommand \doibase [0]{http://dx.doi.org/}%
\providecommand \selectlanguage [0]{\@gobble}%
\providecommand \bibinfo  [0]{\@secondoftwo}%
\providecommand \bibfield  [0]{\@secondoftwo}%
\providecommand \translation [1]{[#1]}%
\providecommand \BibitemOpen [0]{}%
\providecommand \bibitemStop [0]{}%
\providecommand \bibitemNoStop [0]{.\EOS\space}%
\providecommand \EOS [0]{\spacefactor3000\relax}%
\providecommand \BibitemShut  [1]{\csname bibitem#1\endcsname}%
\let\auto@bib@innerbib\@empty
\bibitem [{\citenamefont {Flory}\ and\ \citenamefont {{Rehner,
  Jr.}}(1943{\natexlab{a}})}]{flory-jcp-1943b}%
  \BibitemOpen
  \bibfield  {author} {\bibinfo {author} {\bibfnamefont {P.~J.}\ \bibnamefont
  {Flory}}\ and\ \bibinfo {author} {\bibfnamefont {J.}~\bibnamefont {{Rehner,
  Jr.}}},\ }\bibfield  {title} {\enquote {\bibinfo {title} {Statistical
  mechanics of cross-linked polymer networks {II}. {Swelling}},}\ }\href
  {\doibase 10.1063/1.1723792} {\bibfield  {journal} {\bibinfo  {journal}
  {Journal of Chemical Physics}\ }\textbf {\bibinfo {volume} {11}},\ \bibinfo
  {pages} {521--526} (\bibinfo {year} {1943}{\natexlab{a}})}\BibitemShut
  {NoStop}%
\bibitem [{\citenamefont {Quesada-Perez}\ \emph {et~al.}(2011)\citenamefont
  {Quesada-Perez}, \citenamefont {Maroto-Centeno}, \citenamefont {Forcada},\
  and\ \citenamefont {Hidalgo-Alvarez}}]{quesada-perez-softmatter-2011a}%
  \BibitemOpen
  \bibfield  {author} {\bibinfo {author} {\bibfnamefont {M.}~\bibnamefont
  {Quesada-Perez}}, \bibinfo {author} {\bibfnamefont {J.~A.}\ \bibnamefont
  {Maroto-Centeno}}, \bibinfo {author} {\bibfnamefont {J.}~\bibnamefont
  {Forcada}}, \ and\ \bibinfo {author} {\bibfnamefont {R.}~\bibnamefont
  {Hidalgo-Alvarez}},\ }\bibfield  {title} {\enquote {\bibinfo {title} {Gel
  swelling theories: the classical formalism and recent approaches},}\ }\href
  {\doibase 10.1039/C1SM06031G} {\bibfield  {journal} {\bibinfo  {journal}
  {Soft Matter}\ }\textbf {\bibinfo {volume} {7}},\ \bibinfo {pages}
  {10536--10547} (\bibinfo {year} {2011})}\BibitemShut {NoStop}%
\bibitem [{\citenamefont {Tanaka}\ \emph {et~al.}(1987)\citenamefont {Tanaka},
  \citenamefont {Sun}, \citenamefont {Hirokawa}, \citenamefont {Katayama},
  \citenamefont {Kucera}, \citenamefont {Hirose},\ and\ \citenamefont
  {Amiya}}]{tanaka-n-1987a}%
  \BibitemOpen
  \bibfield  {author} {\bibinfo {author} {\bibfnamefont {T.}~\bibnamefont
  {Tanaka}}, \bibinfo {author} {\bibfnamefont {S.-T.}\ \bibnamefont {Sun}},
  \bibinfo {author} {\bibfnamefont {Y.}~\bibnamefont {Hirokawa}}, \bibinfo
  {author} {\bibfnamefont {S.}~\bibnamefont {Katayama}}, \bibinfo {author}
  {\bibfnamefont {J.}~\bibnamefont {Kucera}}, \bibinfo {author} {\bibfnamefont
  {Y.}~\bibnamefont {Hirose}}, \ and\ \bibinfo {author} {\bibfnamefont
  {T.}~\bibnamefont {Amiya}},\ }\bibfield  {title} {\enquote {\bibinfo {title}
  {Mechanical instability of gels at the phase transition},}\ }\href
  {http://dx.doi.org/10.1038/325796a0} {\bibfield  {journal} {\bibinfo
  {journal} {Nature}\ }\textbf {\bibinfo {volume} {325}},\ \bibinfo {pages}
  {796--798} (\bibinfo {year} {1987})}\BibitemShut {NoStop}%
\bibitem [{\citenamefont {Hong}\ \emph {et~al.}(2008)\citenamefont {Hong},
  \citenamefont {Zhao}, \citenamefont {Zhou},\ and\ \citenamefont
  {Suo}}]{hong-jmps-2008a}%
  \BibitemOpen
  \bibfield  {author} {\bibinfo {author} {\bibfnamefont {W.}~\bibnamefont
  {Hong}}, \bibinfo {author} {\bibfnamefont {X.}~\bibnamefont {Zhao}}, \bibinfo
  {author} {\bibfnamefont {J.}~\bibnamefont {Zhou}}, \ and\ \bibinfo {author}
  {\bibfnamefont {Z.}~\bibnamefont {Suo}},\ }\bibfield  {title} {\enquote
  {\bibinfo {title} {A theory of coupled diffusion and large deformation in
  polymeric gels},}\ }\href {\doibase
  http://dx.doi.org/10.1016/j.jmps.2007.11.010} {\bibfield  {journal} {\bibinfo
   {journal} {Journal of the Mechanics and Physics of Solids}\ }\textbf
  {\bibinfo {volume} {56}},\ \bibinfo {pages} {1779--1793} (\bibinfo {year}
  {2008})}\BibitemShut {NoStop}%
\bibitem [{\citenamefont {Doi}(209)}]{doi-jpsjpn-2009}%
  \BibitemOpen
  \bibfield  {author} {\bibinfo {author} {\bibfnamefont {M.}~\bibnamefont
  {Doi}},\ }\bibfield  {title} {\enquote {\bibinfo {title} {Gel dynamics},}\
  }\href {\doibase http://dx.doi.org/10.1143/JPSJ.78.052001} {\bibfield
  {journal} {\bibinfo  {journal} {Journal of the Physical Society of Japan}\
  }\textbf {\bibinfo {volume} {78}},\ \bibinfo {pages} {052001} (\bibinfo
  {year} {209})}\BibitemShut {NoStop}%
\bibitem [{\citenamefont {Chester}\ and\ \citenamefont
  {Anand}(2010{\natexlab{a}})}]{chester-jmps-2010a}%
  \BibitemOpen
  \bibfield  {author} {\bibinfo {author} {\bibfnamefont {S.~A.}\ \bibnamefont
  {Chester}}\ and\ \bibinfo {author} {\bibfnamefont {L.}~\bibnamefont
  {Anand}},\ }\bibfield  {title} {\enquote {\bibinfo {title} {A coupled theory
  of fluid permeation and large deformations for elastomeric materials},}\
  }\href {\doibase http://dx.doi.org/10.1016/j.jmps.2010.07.020} {\bibfield
  {journal} {\bibinfo  {journal} {Journal of the Mechanics and Physics of
  Solids}\ }\textbf {\bibinfo {volume} {58}},\ \bibinfo {pages} {1879--1906}
  (\bibinfo {year} {2010}{\natexlab{a}})}\BibitemShut {NoStop}%
\bibitem [{\citenamefont {Dervaux}\ \emph {et~al.}(2011)\citenamefont
  {Dervaux}, \citenamefont {Couder}, \citenamefont {Guedeau-Boudeville},\ and\
  \citenamefont {{Ben Amar}}}]{dervaux-prl-2011a}%
  \BibitemOpen
  \bibfield  {author} {\bibinfo {author} {\bibfnamefont {J.}~\bibnamefont
  {Dervaux}}, \bibinfo {author} {\bibfnamefont {Y.}~\bibnamefont {Couder}},
  \bibinfo {author} {\bibfnamefont {M.-A.}\ \bibnamefont {Guedeau-Boudeville}},
  \ and\ \bibinfo {author} {\bibfnamefont {M.}~\bibnamefont {{Ben Amar}}},\
  }\bibfield  {title} {\enquote {\bibinfo {title} {Shape transition in
  artificial tumors: From smooth buckles to singular creases},}\ }\href
  {\doibase 10.1103/PhysRevLett.107.018103} {\bibfield  {journal} {\bibinfo
  {journal} {Physical Review Letters}\ }\textbf {\bibinfo {volume} {107}},\
  \bibinfo {pages} {018103} (\bibinfo {year} {2011})}\BibitemShut {NoStop}%
\bibitem [{\citenamefont {{Barros, Jr.}}\ \emph {et~al.}(2012)\citenamefont
  {{Barros, Jr.}}, \citenamefont {{de Azevedo}},\ and\ \citenamefont
  {Engelsberg}}]{barros-softmatter-2012a}%
  \BibitemOpen
  \bibfield  {author} {\bibinfo {author} {\bibfnamefont {W.}~\bibnamefont
  {{Barros, Jr.}}}, \bibinfo {author} {\bibfnamefont {E.~N.}\ \bibnamefont {{de
  Azevedo}}}, \ and\ \bibinfo {author} {\bibfnamefont {M.}~\bibnamefont
  {Engelsberg}},\ }\bibfield  {title} {\enquote {\bibinfo {title} {Surface
  pattern formation in a swelling gel},}\ }\href {\doibase 10.1039/C2SM25617G}
  {\bibfield  {journal} {\bibinfo  {journal} {Soft Matter}\ }\textbf {\bibinfo
  {volume} {8}},\ \bibinfo {pages} {8511--8516} (\bibinfo {year}
  {2012})}\BibitemShut {NoStop}%
\bibitem [{\citenamefont {Tallinen}\ \emph {et~al.}(2016)\citenamefont
  {Tallinen}, \citenamefont {Chung}, \citenamefont {Rousseau}, \citenamefont
  {Girard}, \citenamefont {Lef{\`{e}}vre},\ and\ \citenamefont
  {Mahadevan}}]{tallinen-natphys-2016}%
  \BibitemOpen
  \bibfield  {author} {\bibinfo {author} {\bibfnamefont {T.}~\bibnamefont
  {Tallinen}}, \bibinfo {author} {\bibfnamefont {J.~Y.}\ \bibnamefont {Chung}},
  \bibinfo {author} {\bibfnamefont {F.}~\bibnamefont {Rousseau}}, \bibinfo
  {author} {\bibfnamefont {N.}~\bibnamefont {Girard}}, \bibinfo {author}
  {\bibfnamefont {J.}~\bibnamefont {Lef{\`{e}}vre}}, \ and\ \bibinfo {author}
  {\bibfnamefont {L.}~\bibnamefont {Mahadevan}},\ }\bibfield  {title} {\enquote
  {\bibinfo {title} {On the growth and form of cortical convolutions},}\ }\href
  {\doibase 10.1038/nphys3632} {\bibfield  {journal} {\bibinfo  {journal}
  {Nature Physics}\ }\textbf {\bibinfo {volume} {12}},\ \bibinfo {pages}
  {588--593} (\bibinfo {year} {2016})}\BibitemShut {NoStop}%
\bibitem [{\citenamefont {Takahashi}\ \emph {et~al.}(2016)\citenamefont
  {Takahashi}, \citenamefont {Ikura}, \citenamefont {King}, \citenamefont
  {Nonoyama}, \citenamefont {Nakajima}, \citenamefont {Kurokawa}, \citenamefont
  {Kuroda}, \citenamefont {Tonegawa},\ and\ \citenamefont
  {Gong}}]{takahashi-softmatter-2016}%
  \BibitemOpen
  \bibfield  {author} {\bibinfo {author} {\bibfnamefont {R.}~\bibnamefont
  {Takahashi}}, \bibinfo {author} {\bibfnamefont {Y.}~\bibnamefont {Ikura}},
  \bibinfo {author} {\bibfnamefont {D.~R.}\ \bibnamefont {King}}, \bibinfo
  {author} {\bibfnamefont {T.}~\bibnamefont {Nonoyama}}, \bibinfo {author}
  {\bibfnamefont {T.}~\bibnamefont {Nakajima}}, \bibinfo {author}
  {\bibfnamefont {T.}~\bibnamefont {Kurokawa}}, \bibinfo {author}
  {\bibfnamefont {H.}~\bibnamefont {Kuroda}}, \bibinfo {author} {\bibfnamefont
  {Y.}~\bibnamefont {Tonegawa}}, \ and\ \bibinfo {author} {\bibfnamefont
  {J.~P.}\ \bibnamefont {Gong}},\ }\bibfield  {title} {\enquote {\bibinfo
  {title} {Coupled instabilities of surface crease and bulk bending during fast
  free swelling of hydrogels},}\ }\href {\doibase 10.1039/C6SM00578K}
  {\bibfield  {journal} {\bibinfo  {journal} {Soft Matter}\ }\textbf {\bibinfo
  {volume} {12}},\ \bibinfo {pages} {5081--5088} (\bibinfo {year}
  {2016})}\BibitemShut {NoStop}%
\bibitem [{\citenamefont {Zohuriaan-Mehr}\ \emph {et~al.}(2010)\citenamefont
  {Zohuriaan-Mehr}, \citenamefont {Omidian}, \citenamefont {Doroudiani},\ and\
  \citenamefont {Kabiri}}]{zohuriaan-mehr-jmatsci-2010}%
  \BibitemOpen
  \bibfield  {author} {\bibinfo {author} {\bibfnamefont {M.~J.}\ \bibnamefont
  {Zohuriaan-Mehr}}, \bibinfo {author} {\bibfnamefont {H.}~\bibnamefont
  {Omidian}}, \bibinfo {author} {\bibfnamefont {S.}~\bibnamefont {Doroudiani}},
  \ and\ \bibinfo {author} {\bibfnamefont {K.}~\bibnamefont {Kabiri}},\
  }\bibfield  {title} {\enquote {\bibinfo {title} {Advances in non-hygienic
  applications of superabsorbent hydrogel materials},}\ }\href {\doibase
  10.1007/s10853-010-4780-1} {\bibfield  {journal} {\bibinfo  {journal}
  {Journal of Materials Science}\ }\textbf {\bibinfo {volume} {45}},\ \bibinfo
  {pages} {5711---5735} (\bibinfo {year} {2010})}\BibitemShut {NoStop}%
\bibitem [{\citenamefont {Peppas}\ \emph {et~al.}(2000)\citenamefont {Peppas},
  \citenamefont {Huang}, \citenamefont {Torres-Lugo}, \citenamefont {Ward},\
  and\ \citenamefont {Zhang}}]{peppas-annrevbiomedeng-2000}%
  \BibitemOpen
  \bibfield  {author} {\bibinfo {author} {\bibfnamefont {N.~A.}\ \bibnamefont
  {Peppas}}, \bibinfo {author} {\bibfnamefont {Y.}~\bibnamefont {Huang}},
  \bibinfo {author} {\bibfnamefont {M.}~\bibnamefont {Torres-Lugo}}, \bibinfo
  {author} {\bibfnamefont {J.~H.}\ \bibnamefont {Ward}}, \ and\ \bibinfo
  {author} {\bibfnamefont {J.}~\bibnamefont {Zhang}},\ }\bibfield  {title}
  {\enquote {\bibinfo {title} {Physicochemical foundations and structural
  design of hydrogels in medicine and biology},}\ }\href {\doibase
  10.1146/annurev.bioeng.2.1.9} {\bibfield  {journal} {\bibinfo  {journal}
  {Annual Review of Biomedical Engineering}\ }\textbf {\bibinfo {volume} {2}},\
  \bibinfo {pages} {9--29} (\bibinfo {year} {2000})}\BibitemShut {NoStop}%
\bibitem [{\citenamefont {Cal{\'o}}\ and\ \citenamefont
  {Khutoryanskiy}(2015)}]{calo-eurpolymj-2015}%
  \BibitemOpen
  \bibfield  {author} {\bibinfo {author} {\bibfnamefont {E.}~\bibnamefont
  {Cal{\'o}}}\ and\ \bibinfo {author} {\bibfnamefont {V.~V.}\ \bibnamefont
  {Khutoryanskiy}},\ }\bibfield  {title} {\enquote {\bibinfo {title}
  {Biomedical applications of hydrogels: {A} review of patents and commercial
  products},}\ }\href {\doibase 10.1016/j.eurpolymj.2014.11.024} {\bibfield
  {journal} {\bibinfo  {journal} {European Polymer Journal}\ }\textbf {\bibinfo
  {volume} {65}},\ \bibinfo {pages} {252--267} (\bibinfo {year}
  {2015})}\BibitemShut {NoStop}%
\bibitem [{\citenamefont {Langer}(1998)}]{langer-nature-1998}%
  \BibitemOpen
  \bibfield  {author} {\bibinfo {author} {\bibfnamefont {R.}~\bibnamefont
  {Langer}},\ }\bibfield  {title} {\enquote {\bibinfo {title} {Drug delivery
  and targeting},}\ }\bibfield  {booktitle} {\emph {\bibinfo {booktitle}
  {Therapeutic Horizons (supplement to issue 6679)}},\ }\href {\doibase
  http://www.nature.com/nature/supplements/collections/therapeutichorizons/pdf/392005a.pdf}
  {\bibfield  {journal} {\bibinfo  {journal} {Nature}\ }\textbf {\bibinfo
  {volume} {392}},\ \bibinfo {pages} {5--10} (\bibinfo {year} {1998})}\BibitemShut
  {NoStop}%
\bibitem [{\citenamefont {Cohen}\ and\ \citenamefont
  {Erneux}(1988)}]{cohen-sjam-1988a}%
  \BibitemOpen
  \bibfield  {author} {\bibinfo {author} {\bibfnamefont {D.}~\bibnamefont
  {Cohen}}\ and\ \bibinfo {author} {\bibfnamefont {T.}~\bibnamefont {Erneux}},\
  }\bibfield  {title} {\enquote {\bibinfo {title} {Free boundary problems in
  controlled release pharmaceuticals: {II}. {Swelling}-controlled release},}\
  }\href {\doibase 10.1137/0148090} {\bibfield  {journal} {\bibinfo  {journal}
  {SIAM Journal on Applied Mathematics}\ }\textbf {\bibinfo {volume} {48}},\
  \bibinfo {pages} {1466--1474} (\bibinfo {year} {1988})}\BibitemShut {NoStop}%
\bibitem [{\citenamefont {Harmon}\ \emph {et~al.}(2003)\citenamefont {Harmon},
  \citenamefont {Tang},\ and\ \citenamefont {Frank}}]{harmon-polymer-2003}%
  \BibitemOpen
  \bibfield  {author} {\bibinfo {author} {\bibfnamefont {M.~E.}\ \bibnamefont
  {Harmon}}, \bibinfo {author} {\bibfnamefont {M.}~\bibnamefont {Tang}}, \ and\
  \bibinfo {author} {\bibfnamefont {C.~W.}\ \bibnamefont {Frank}},\ }\bibfield
  {title} {\enquote {\bibinfo {title} {A microfluidic actuator based on
  thermoresponsive hydrogels},}\ }\href {\doibase
  10.1016/S0032-3861(03)00463-4} {\bibfield  {journal} {\bibinfo  {journal}
  {Polymer}\ }\textbf {\bibinfo {volume} {44}},\ \bibinfo {pages} {4547--4556}
  (\bibinfo {year} {2003})}\BibitemShut {NoStop}%
\bibitem [{\citenamefont {Deligkaris}\ \emph {et~al.}(2010)\citenamefont
  {Deligkaris}, \citenamefont {Tadele}, \citenamefont {Olthuis},\ and\
  \citenamefont {{van den Berg}}}]{deligkaris-snb-2010}%
  \BibitemOpen
  \bibfield  {author} {\bibinfo {author} {\bibfnamefont {K.}~\bibnamefont
  {Deligkaris}}, \bibinfo {author} {\bibfnamefont {T.~S.}\ \bibnamefont
  {Tadele}}, \bibinfo {author} {\bibfnamefont {W.}~\bibnamefont {Olthuis}}, \
  and\ \bibinfo {author} {\bibfnamefont {A.}~\bibnamefont {{van den Berg}}},\
  }\bibfield  {title} {\enquote {\bibinfo {title} {Hydrogel-based devices for
  biomedical applications},}\ }\href {\doibase 10.1016/j.snb.2010.03.083}
  {\bibfield  {journal} {\bibinfo  {journal} {Sensors and Actuators B:
  Chemical}\ }\textbf {\bibinfo {volume} {147}},\ \bibinfo {pages} {765---774}
  (\bibinfo {year} {2010})}\BibitemShut {NoStop}%
\bibitem [{\citenamefont {Holmes}\ \emph {et~al.}(2011)\citenamefont {Holmes},
  \citenamefont {Roch\'{e}}, \citenamefont {Sinha},\ and\ \citenamefont
  {Stone}}]{holmes-softmatter-2011}%
  \BibitemOpen
  \bibfield  {author} {\bibinfo {author} {\bibfnamefont {D.~P.}\ \bibnamefont
  {Holmes}}, \bibinfo {author} {\bibfnamefont {M.}~\bibnamefont {Roch\'{e}}},
  \bibinfo {author} {\bibfnamefont {T.}~\bibnamefont {Sinha}}, \ and\ \bibinfo
  {author} {\bibfnamefont {H.~A.}\ \bibnamefont {Stone}},\ }\bibfield  {title}
  {\enquote {\bibinfo {title} {Bending and twisting of soft materials by
  non-homogeneous swelling},}\ }\href {\doibase 10.1039/c0sm01492c} {\bibfield
  {journal} {\bibinfo  {journal} {Soft Matter}\ }\textbf {\bibinfo {volume}
  {7}},\ \bibinfo {pages} {5188--5193} (\bibinfo {year} {2011})}\BibitemShut
  {NoStop}%
\bibitem [{\citenamefont {Mukhopadhyay}\ and\ \citenamefont
  {Peixinho}(2011)}]{mukhopadhyay-pre-2011a}%
  \BibitemOpen
  \bibfield  {author} {\bibinfo {author} {\bibfnamefont {S.}~\bibnamefont
  {Mukhopadhyay}}\ and\ \bibinfo {author} {\bibfnamefont {J.}~\bibnamefont
  {Peixinho}},\ }\bibfield  {title} {\enquote {\bibinfo {title} {Packings of
  deformable spheres},}\ }\href {\doibase 10.1103/PhysRevE.84.011302}
  {\bibfield  {journal} {\bibinfo  {journal} {Physical Review E}\ }\textbf
  {\bibinfo {volume} {84}},\ \bibinfo {pages} {011302} (\bibinfo {year}
  {2011})}\BibitemShut {NoStop}%
\bibitem [{\citenamefont {MacMinn}\ \emph {et~al.}(2015)\citenamefont
  {MacMinn}, \citenamefont {Dufresne},\ and\ \citenamefont
  {Wettlaufer}}]{macminn-prx-2015a}%
  \BibitemOpen
  \bibfield  {author} {\bibinfo {author} {\bibfnamefont {C.~W.}\ \bibnamefont
  {MacMinn}}, \bibinfo {author} {\bibfnamefont {E.~R.}\ \bibnamefont
  {Dufresne}}, \ and\ \bibinfo {author} {\bibfnamefont {J.~S.}\ \bibnamefont
  {Wettlaufer}},\ }\bibfield  {title} {\enquote {\bibinfo {title} {Fluid-driven
  deformation of a soft granular material},}\ }\href {\doibase
  10.1103/PhysRevX.5.011020} {\bibfield  {journal} {\bibinfo  {journal}
  {Physical Review X}\ }\textbf {\bibinfo {volume} {5}},\ \bibinfo {pages}
  {011020} (\bibinfo {year} {2015})}\BibitemShut {NoStop}%
\bibitem [{\citenamefont {Brodu}\ \emph {et~al.}(2015)\citenamefont {Brodu},
  \citenamefont {Dijksman},\ and\ \citenamefont
  {Behringer}}]{brodu-natcomm-2015a}%
  \BibitemOpen
  \bibfield  {author} {\bibinfo {author} {\bibfnamefont {N.}~\bibnamefont
  {Brodu}}, \bibinfo {author} {\bibfnamefont {J.~A.}\ \bibnamefont {Dijksman}},
  \ and\ \bibinfo {author} {\bibfnamefont {R.~P.}\ \bibnamefont {Behringer}},\
  }\bibfield  {title} {\enquote {\bibinfo {title} {Spanning the scales of
  granular materials through microscopic force imaging},}\ }\href {\doibase
  10.1038/ncomms7361} {\bibfield  {journal} {\bibinfo  {journal} {Nature
  Communications}\ }\textbf {\bibinfo {volume} {6}},\ \bibinfo {pages} {6361}
  (\bibinfo {year} {2015})}\BibitemShut {NoStop}%
\bibitem [{\citenamefont {Cai}\ and\ \citenamefont
  {Suo}(2012)}]{cai-epl-2012a}%
  \BibitemOpen
  \bibfield  {author} {\bibinfo {author} {\bibfnamefont {S.}~\bibnamefont
  {Cai}}\ and\ \bibinfo {author} {\bibfnamefont {Z.}~\bibnamefont {Suo}},\
  }\bibfield  {title} {\enquote {\bibinfo {title} {Equations of state for ideal
  elastomeric gels},}\ }\href {http://stacks.iop.org/0295-5075/97/i=3/a=34009}
  {\bibfield  {journal} {\bibinfo  {journal} {EPL (Europhysics Letters)}\
  }\textbf {\bibinfo {volume} {97}},\ \bibinfo {pages} {34009} (\bibinfo {year}
  {2012})}\BibitemShut {NoStop}%
\bibitem [{\citenamefont {Li}\ \emph {et~al.}(2012)\citenamefont {Li},
  \citenamefont {Hu}, \citenamefont {Vlassak},\ and\ \citenamefont
  {Suo}}]{li-softmatter-2012a}%
  \BibitemOpen
  \bibfield  {author} {\bibinfo {author} {\bibfnamefont {J.}~\bibnamefont
  {Li}}, \bibinfo {author} {\bibfnamefont {Y.}~\bibnamefont {Hu}}, \bibinfo
  {author} {\bibfnamefont {J.~J.}\ \bibnamefont {Vlassak}}, \ and\ \bibinfo
  {author} {\bibfnamefont {Z.}~\bibnamefont {Suo}},\ }\bibfield  {title}
  {\enquote {\bibinfo {title} {Experimental determination of equations of state
  for ideal elastomeric gels},}\ }\href {\doibase 10.1039/C2SM25437A}
  {\bibfield  {journal} {\bibinfo  {journal} {Soft Matter}\ }\textbf {\bibinfo
  {volume} {8}},\ \bibinfo {pages} {8121--8128} (\bibinfo {year}
  {2012})}\BibitemShut {NoStop}%
\bibitem [{\citenamefont {Tanaka}\ \emph {et~al.}(1973)\citenamefont {Tanaka},
  \citenamefont {Hocker},\ and\ \citenamefont
  {Benedek}}]{tanaka-jchemphys-1973}%
  \BibitemOpen
  \bibfield  {author} {\bibinfo {author} {\bibfnamefont {T.}~\bibnamefont
  {Tanaka}}, \bibinfo {author} {\bibfnamefont {L.~O.}\ \bibnamefont {Hocker}},
  \ and\ \bibinfo {author} {\bibfnamefont {G.~B.}\ \bibnamefont {Benedek}},\
  }\bibfield  {title} {\enquote {\bibinfo {title} {Spectrum of light scattered
  from a viscoelastic gel},}\ }\href {\doibase 10.1063/1.1680734} {\bibfield
  {journal} {\bibinfo  {journal} {Journal of Chemical Physics}\ }\textbf
  {\bibinfo {volume} {59}},\ \bibinfo {pages} {5151} (\bibinfo {year}
  {1973})}\BibitemShut {NoStop}%
\bibitem [{\citenamefont {Tanaka}\ and\ \citenamefont
  {Fillmore}(1979)}]{tanaka-jchemphys-1979}%
  \BibitemOpen
  \bibfield  {author} {\bibinfo {author} {\bibfnamefont {T.}~\bibnamefont
  {Tanaka}}\ and\ \bibinfo {author} {\bibfnamefont {D.~J.}\ \bibnamefont
  {Fillmore}},\ }\bibfield  {title} {\enquote {\bibinfo {title} {Kinetics of
  swelling of gels},}\ }\href {\doibase 10.1063/1.437602} {\bibfield  {journal}
  {\bibinfo  {journal} {Journal of Chemical Physics}\ }\textbf {\bibinfo
  {volume} {70}},\ \bibinfo {pages} {1214} (\bibinfo {year}
  {1979})}\BibitemShut {NoStop}%
\bibitem [{\citenamefont {Johnson}(1982)}]{johnson-jchemphys-1982}%
  \BibitemOpen
  \bibfield  {author} {\bibinfo {author} {\bibfnamefont {D.~L.}\ \bibnamefont
  {Johnson}},\ }\bibfield  {title} {\enquote {\bibinfo {title} {Elastodynamics
  of gels},}\ }\href {\doibase 10.1063/1.443934} {\bibfield  {journal}
  {\bibinfo  {journal} {Journal of Chemical Physics}\ }\textbf {\bibinfo
  {volume} {77}},\ \bibinfo {pages} {1531} (\bibinfo {year}
  {1982})}\BibitemShut {NoStop}%
\bibitem [{\citenamefont {Scherer}(1989)}]{scherer-jnoncrystsolids-1989}%
  \BibitemOpen
  \bibfield  {author} {\bibinfo {author} {\bibfnamefont {G.~W.}\ \bibnamefont
  {Scherer}},\ }\bibfield  {title} {\enquote {\bibinfo {title} {Measurement of
  permeability {I}. {Theory}},}\ }\href {\doibase 10.1016/0022-3093(89)90001-X}
  {\bibfield  {journal} {\bibinfo  {journal} {Journal of Non-Crystalline
  Solids}\ }\textbf {\bibinfo {volume} {113}},\ \bibinfo {pages} {107--118}
  (\bibinfo {year} {1989})}\BibitemShut {NoStop}%
\bibitem [{\citenamefont {Hui}\ and\ \citenamefont
  {Muralidharan}(2005)}]{hui-jchemphys-2005}%
  \BibitemOpen
  \bibfield  {author} {\bibinfo {author} {\bibfnamefont {C.-Y.}\ \bibnamefont
  {Hui}}\ and\ \bibinfo {author} {\bibfnamefont {V.}~\bibnamefont
  {Muralidharan}},\ }\bibfield  {title} {\enquote {\bibinfo {title} {Gel
  mechanics: {A} comparison of the theories of {Biot}, {Tanaka}, {Hocker}, and
  {Benedek}},}\ }\href {\doibase 10.1063/1.2061987} {\bibfield  {journal}
  {\bibinfo  {journal} {Journal of Chemical Physics}\ }\textbf {\bibinfo
  {volume} {123}},\ \bibinfo {pages} {154905} (\bibinfo {year}
  {2005})}\BibitemShut {NoStop}%
\bibitem [{\citenamefont {Wahrmund}\ \emph {et~al.}(2009)\citenamefont
  {Wahrmund}, \citenamefont {Kim}, \citenamefont {Chu}, \citenamefont {Wang},
  \citenamefont {Li}, \citenamefont {Fernandez-Nieves}, \citenamefont {Weitz},
  \citenamefont {Krokhin},\ and\ \citenamefont {Hu}}]{wahrmund-macromol-2009a}%
  \BibitemOpen
  \bibfield  {author} {\bibinfo {author} {\bibfnamefont {J.}~\bibnamefont
  {Wahrmund}}, \bibinfo {author} {\bibfnamefont {J.-W.}\ \bibnamefont {Kim}},
  \bibinfo {author} {\bibfnamefont {L.-Y.}\ \bibnamefont {Chu}}, \bibinfo
  {author} {\bibfnamefont {C.}~\bibnamefont {Wang}}, \bibinfo {author}
  {\bibfnamefont {Y.}~\bibnamefont {Li}}, \bibinfo {author} {\bibfnamefont
  {A.}~\bibnamefont {Fernandez-Nieves}}, \bibinfo {author} {\bibfnamefont
  {D.~A.}\ \bibnamefont {Weitz}}, \bibinfo {author} {\bibfnamefont
  {A.}~\bibnamefont {Krokhin}}, \ and\ \bibinfo {author} {\bibfnamefont
  {Z.}~\bibnamefont {Hu}},\ }\bibfield  {title} {\enquote {\bibinfo {title}
  {Swelling kinetics of a microgel shell},}\ }\href {\doibase
  10.1021/ma901362p} {\bibfield  {journal} {\bibinfo  {journal}
  {Macromolecules}\ }\textbf {\bibinfo {volume} {42}},\ \bibinfo {pages}
  {9357--9365} (\bibinfo {year} {2009})}\BibitemShut {NoStop}%
\bibitem [{\citenamefont {Yoon}\ \emph {et~al.}(2010)\citenamefont {Yoon},
  \citenamefont {Cai}, \citenamefont {Suo},\ and\ \citenamefont
  {Hayward}}]{yoon-softmatter-2010a}%
  \BibitemOpen
  \bibfield  {author} {\bibinfo {author} {\bibfnamefont {J.}~\bibnamefont
  {Yoon}}, \bibinfo {author} {\bibfnamefont {S.}~\bibnamefont {Cai}}, \bibinfo
  {author} {\bibfnamefont {Z.}~\bibnamefont {Suo}}, \ and\ \bibinfo {author}
  {\bibfnamefont {R.~C.}\ \bibnamefont {Hayward}},\ }\bibfield  {title}
  {\enquote {\bibinfo {title} {Poroelastic swelling kinetics of thin hydrogel
  layers: comparison of theory and experiment},}\ }\href {\doibase
  10.1039/C0SM00434K} {\bibfield  {journal} {\bibinfo  {journal} {Soft Matter}\
  }\textbf {\bibinfo {volume} {6}},\ \bibinfo {pages} {6004--6012} (\bibinfo
  {year} {2010})}\BibitemShut {NoStop}%
\bibitem [{\citenamefont {Duda}\ \emph {et~al.}(2010)\citenamefont {Duda},
  \citenamefont {Souza},\ and\ \citenamefont {Fried}}]{duda-jmps-2010}%
  \BibitemOpen
  \bibfield  {author} {\bibinfo {author} {\bibfnamefont {F.~P.}\ \bibnamefont
  {Duda}}, \bibinfo {author} {\bibfnamefont {A.~C.}\ \bibnamefont {Souza}}, \
  and\ \bibinfo {author} {\bibfnamefont {E.}~\bibnamefont {Fried}},\ }\bibfield
   {title} {\enquote {\bibinfo {title} {A theory for species migration in a
  finitely strained solid with application to polymer network swelling},}\
  }\href {\doibase 10.1016/j.jmps.2010.01.009} {\bibfield  {journal} {\bibinfo
  {journal} {Journal of the Mechanical and Physics of Solids}\ }\textbf
  {\bibinfo {volume} {58}},\ \bibinfo {pages} {515--529} (\bibinfo {year}
  {2010})}\BibitemShut {NoStop}%
\bibitem [{\citenamefont {Bouklas}\ \emph {et~al.}(2015)\citenamefont
  {Bouklas}, \citenamefont {Landis},\ and\ \citenamefont
  {Huang}}]{bouklas-jmps-2015a}%
  \BibitemOpen
  \bibfield  {author} {\bibinfo {author} {\bibfnamefont {N.}~\bibnamefont
  {Bouklas}}, \bibinfo {author} {\bibfnamefont {C.~M.}\ \bibnamefont {Landis}},
  \ and\ \bibinfo {author} {\bibfnamefont {R.}~\bibnamefont {Huang}},\
  }\bibfield  {title} {\enquote {\bibinfo {title} {A nonlinear, transient
  finite element method for coupled solvent diffusion and large deformation of
  hydrogels},}\ }\href {\doibase http://dx.doi.org/10.1016/j.jmps.2015.03.004}
  {\bibfield  {journal} {\bibinfo  {journal} {Journal of the Mechanics and
  Physics of Solids}\ }\textbf {\bibinfo {volume} {79}},\ \bibinfo {pages}
  {21--43} (\bibinfo {year} {2015})}\BibitemShut {NoStop}%
\bibitem [{\citenamefont {Engelsberg}\ and\ \citenamefont {{Barros,
  Jr.}}(2013)}]{engelsberg-pre-2013a}%
  \BibitemOpen
  \bibfield  {author} {\bibinfo {author} {\bibfnamefont {M.}~\bibnamefont
  {Engelsberg}}\ and\ \bibinfo {author} {\bibfnamefont {W.}~\bibnamefont
  {{Barros, Jr.}}},\ }\bibfield  {title} {\enquote {\bibinfo {title}
  {Free-evolution kinetics in a high-swelling polymeric hydrogel},}\ }\href
  {\doibase 10.1103/PhysRevE.88.062602} {\bibfield  {journal} {\bibinfo
  {journal} {Phys. Rev. E}\ }\textbf {\bibinfo {volume} {88}},\ \bibinfo
  {pages} {062602} (\bibinfo {year} {2013})}\BibitemShut {NoStop}%
\bibitem [{\citenamefont {MacMinn}\ \emph {et~al.}(2016)\citenamefont
  {MacMinn}, \citenamefont {Dufresne},\ and\ \citenamefont
  {Wettlaufer}}]{macminn-prapplied-2016}%
  \BibitemOpen
  \bibfield  {author} {\bibinfo {author} {\bibfnamefont {C.~W.}\ \bibnamefont
  {MacMinn}}, \bibinfo {author} {\bibfnamefont {E.~R.}\ \bibnamefont
  {Dufresne}}, \ and\ \bibinfo {author} {\bibfnamefont {J.~S.}\ \bibnamefont
  {Wettlaufer}},\ }\bibfield  {title} {\enquote {\bibinfo {title} {Large
  deformations of a soft porous material},}\ }\href@noop {} {\bibfield
  {journal} {\bibinfo  {journal} {Physical Review Applied}\ }\textbf {\bibinfo
  {volume} {5}},\ \bibinfo {pages} {044020} (\bibinfo {year}
  {2016})}\BibitemShut {NoStop}%
\bibitem [{\citenamefont {Flory}\ and\ \citenamefont {{Rehner,
  Jr.}}(1943{\natexlab{b}})}]{flory-jcp-1943a}%
  \BibitemOpen
  \bibfield  {author} {\bibinfo {author} {\bibfnamefont {P.~J.}\ \bibnamefont
  {Flory}}\ and\ \bibinfo {author} {\bibfnamefont {J.}~\bibnamefont {{Rehner,
  Jr.}}},\ }\bibfield  {title} {\enquote {\bibinfo {title} {Statistical
  mechanics of cross-linked polymer networks {I}. {Rubberlike} elasticity},}\
  }\href {\doibase 10.1063/1.1723791} {\bibfield  {journal} {\bibinfo
  {journal} {Journal of Chemical Physics}\ }\textbf {\bibinfo {volume} {11}},\
  \bibinfo {pages} {512--520} (\bibinfo {year}
  {1943}{\natexlab{b}})}\BibitemShut {NoStop}%
\bibitem [{\citenamefont {Flory}(1942)}]{flory-jchemphys-1942}%
  \BibitemOpen
  \bibfield  {author} {\bibinfo {author} {\bibfnamefont {P.~J.}\ \bibnamefont
  {Flory}},\ }\bibfield  {title} {\enquote {\bibinfo {title} {Thermodynamics of
  high polymer solutions},}\ }\href {\doibase 10.1063/1.1723621} {\bibfield
  {journal} {\bibinfo  {journal} {Journal of Chemical Physics}\ }\textbf
  {\bibinfo {volume} {10}},\ \bibinfo {pages} {51} (\bibinfo {year}
  {1942})}\BibitemShut {NoStop}%
\bibitem [{\citenamefont {Huggins}(1942)}]{huggins-jphyschem-1942}%
  \BibitemOpen
  \bibfield  {author} {\bibinfo {author} {\bibfnamefont {M.~L.}\ \bibnamefont
  {Huggins}},\ }\bibfield  {title} {\enquote {\bibinfo {title} {Some properties
  of solutions of long-chain compounds},}\ }\href {\doibase
  10.1021/j150415a018} {\bibfield  {journal} {\bibinfo  {journal} {J.
  Phys. Chem.}\ }\textbf {\bibinfo {volume} {46}},\ \bibinfo {pages}
  {151--158} (\bibinfo {year} {1942})}\BibitemShut {NoStop}%
\bibitem [{\citenamefont {Boyce}\ and\ \citenamefont
  {Arruda}(2000)}]{boyce-rct-2000a}%
  \BibitemOpen
  \bibfield  {author} {\bibinfo {author} {\bibfnamefont {M.~C.}\ \bibnamefont
  {Boyce}}\ and\ \bibinfo {author} {\bibfnamefont {E.~M.}\ \bibnamefont
  {Arruda}},\ }\bibfield  {title} {\enquote {\bibinfo {title} {Constitutive
  models of rubber elasticity: A review},}\ }\bibfield  {booktitle} {\emph
  {\bibinfo {booktitle} {Rubber Chemistry and Technology}},\ }\href {\doibase
  10.5254/1.3547602} {\bibfield  {journal} {\bibinfo  {journal} {Rubber
  Chemistry and Technology}\ }\textbf {\bibinfo {volume} {73}},\ \bibinfo
  {pages} {504--523} (\bibinfo {year} {2000})}\BibitemShut {NoStop}%
\bibitem [{\citenamefont {Biot}(1941)}]{biot-jap-1941a}%
  \BibitemOpen
  \bibfield  {author} {\bibinfo {author} {\bibfnamefont {M.~A.}\ \bibnamefont
  {Biot}},\ }\bibfield  {title} {\enquote {\bibinfo {title} {General theory of
  three-dimensional consolidation},}\ }\href {\doibase
  http://dx.doi.org/10.1063/1.1712886} {\bibfield  {journal} {\bibinfo
  {journal} {Journal of Applied Physics}\ }\textbf {\bibinfo {volume} {12}},\
  \bibinfo {pages} {155--164} (\bibinfo {year} {1941})}\BibitemShut {NoStop}%
\bibitem [{\citenamefont {Chester}\ and\ \citenamefont
  {Anand}(2010{\natexlab{b}})}]{chester-jmechphyssolids-2010}%
  \BibitemOpen
  \bibfield  {author} {\bibinfo {author} {\bibfnamefont {S.~A.}\ \bibnamefont
  {Chester}}\ and\ \bibinfo {author} {\bibfnamefont {L.}~\bibnamefont
  {Anand}},\ }\bibfield  {title} {\enquote {\bibinfo {title} {A coupled theory
  of fluid permeation and large deformations for elastomeric materials},}\
  }\href {\doibase 10.1016/j.jmps.2010.07.020} {\bibfield  {journal} {\bibinfo
  {journal} {Journal of the Mechanics and Physics of Solids}\ }\textbf
  {\bibinfo {volume} {58}},\ \bibinfo {pages} {1879--1906} (\bibinfo {year}
  {2010}{\natexlab{b}})}\BibitemShut {NoStop}%
\bibitem [{\citenamefont {Peppin}\ \emph {et~al.}(2005)\citenamefont {Peppin},
  \citenamefont {Elliott},\ and\ \citenamefont
  {Worster}}]{peppin-physfluids-2005a}%
  \BibitemOpen
  \bibfield  {author} {\bibinfo {author} {\bibfnamefont {S.~S.~L.}\
  \bibnamefont {Peppin}}, \bibinfo {author} {\bibfnamefont {J.~A.~W.}\
  \bibnamefont {Elliott}}, \ and\ \bibinfo {author} {\bibfnamefont {M.~G.}\
  \bibnamefont {Worster}},\ }\bibfield  {title} {\enquote {\bibinfo {title}
  {Pressure and relative motion in colloidal suspensions},}\ }\href {\doibase
  http://dx.doi.org/10.1063/1.1915027} {\bibfield  {journal} {\bibinfo
  {journal} {Physics of Fluids}\ }\textbf {\bibinfo {volume} {17}},\ \bibinfo
  {pages} {053301} (\bibinfo {year} {2005})}\BibitemShut {NoStop}%
\bibitem [{\citenamefont {Tokita}\ and\ \citenamefont
  {Tanaka}(1991)}]{tokita-jcp-1991a}%
  \BibitemOpen
  \bibfield  {author} {\bibinfo {author} {\bibfnamefont {M.}~\bibnamefont
  {Tokita}}\ and\ \bibinfo {author} {\bibfnamefont {T.}~\bibnamefont
  {Tanaka}},\ }\bibfield  {title} {\enquote {\bibinfo {title} {Friction
  coefficient of polymer networks of gels},}\ }\href {\doibase
  10.1063/1.461729} {\bibfield  {journal} {\bibinfo  {journal} {J.
  Chem. Phys.}\ }\textbf {\bibinfo {volume} {95}},\ \bibinfo {pages}
  {4613--4619} (\bibinfo {year} {1991})}\BibitemShut {NoStop}%
\bibitem [{\citenamefont {Grattoni}\ \emph {et~al.}(2001)\citenamefont
  {Grattoni}, \citenamefont {Al-Sharji}, \citenamefont {Yang}, \citenamefont
  {Muggeridge},\ and\ \citenamefont {Zimmerman}}]{grattoni-jcis-2001a}%
  \BibitemOpen
  \bibfield  {author} {\bibinfo {author} {\bibfnamefont {Carlos~A.}\
  \bibnamefont {Grattoni}}, \bibinfo {author} {\bibfnamefont {Hamed~H.}\
  \bibnamefont {Al-Sharji}}, \bibinfo {author} {\bibfnamefont {Canghu}\
  \bibnamefont {Yang}}, \bibinfo {author} {\bibfnamefont {Ann~H.}\ \bibnamefont
  {Muggeridge}}, \ and\ \bibinfo {author} {\bibfnamefont {Robert~W.}\
  \bibnamefont {Zimmerman}},\ }\bibfield  {title} {\enquote {\bibinfo {title}
  {Rheology and permeability of crosslinked polyacrylamide gel},}\ }\href
  {\doibase http://dx.doi.org/10.1006/jcis.2001.7633} {\bibfield  {journal}
  {\bibinfo  {journal} {Journal of Colloid and Interface Science}\ }\textbf
  {\bibinfo {volume} {240}},\ \bibinfo {pages} {601--607} (\bibinfo {year}
  {2001})}\BibitemShut {NoStop}%
\bibitem [{\citenamefont {Matsuo}\ and\ \citenamefont
  {Tanaka}(1988)}]{matsuo-jcp-1988a}%
  \BibitemOpen
  \bibfield  {author} {\bibinfo {author} {\bibfnamefont {E.~S.}\ \bibnamefont
  {Matsuo}}\ and\ \bibinfo {author} {\bibfnamefont {T.}~\bibnamefont
  {Tanaka}},\ }\bibfield  {title} {\enquote {\bibinfo {title} {Kinetics of
  discontinuous volume--phase transition of gels},}\ }\href {\doibase
  http://dx.doi.org/10.1063/1.455115} {\bibfield  {journal} {\bibinfo
  {journal} {The Journal of Chemical Physics}\ }\textbf {\bibinfo {volume}
  {89}},\ \bibinfo {pages} {1695--1703} (\bibinfo {year} {1988})}\BibitemShut
  {NoStop}%
\bibitem [{\citenamefont {Eden}(1961)}]{eden-proceedings-1961a}%
  \BibitemOpen
  \bibfield  {author} {\bibinfo {author} {\bibfnamefont {M.}~\bibnamefont
  {Eden}},\ }\bibfield  {title} {\enquote {\bibinfo {title} {A two-dimensional
  growth process},}\ }in\ \href
  {http://projecteuclid.org/euclid.bsmsp/1200512888} {\emph {\bibinfo
  {booktitle} {Proceedings of the Fourth Berkeley Symposium on Mathematical
  Statistics and Probability, Volume 4: Contributions to Biology and Problems
  of Medicine}}}\ (\bibinfo  {publisher} {University of California Press},\
  \bibinfo {address} {Berkeley, Calif.},\ \bibinfo {year} {1961})\ pp.\
  \bibinfo {pages} {223--239}\BibitemShut {NoStop}%
\bibitem [{\citenamefont {Kardar}\ \emph {et~al.}(1986)\citenamefont {Kardar},
  \citenamefont {Parisi},\ and\ \citenamefont {Zhang}}]{kardar-prl-1986a}%
  \BibitemOpen
  \bibfield  {author} {\bibinfo {author} {\bibfnamefont {M.}~\bibnamefont
  {Kardar}}, \bibinfo {author} {\bibfnamefont {G.}~\bibnamefont {Parisi}}, \
  and\ \bibinfo {author} {\bibfnamefont {Y.-C.}\ \bibnamefont {Zhang}},\
  }\bibfield  {title} {\enquote {\bibinfo {title} {Dynamic scaling of growing
  interfaces},}\ }\href {\doibase 10.1103/PhysRevLett.56.889} {\bibfield
  {journal} {\bibinfo  {journal} {Phys. Rev. Lett.}\ }\textbf {\bibinfo
  {volume} {56}},\ \bibinfo {pages} {889--892} (\bibinfo {year}
  {1986})}\BibitemShut {NoStop}%
\bibitem [{\citenamefont {Family}(1990)}]{family-physicaa-1990a}%
  \BibitemOpen
  \bibfield  {author} {\bibinfo {author} {\bibfnamefont {F.}~\bibnamefont
  {Family}},\ }\bibfield  {title} {\enquote {\bibinfo {title} {Dynamic scaling
  and phase transitions in interface growth},}\ }\href@noop {} {\bibfield
  {journal} {\bibinfo  {journal} {Physica A: Statistical Mechanics and its
  Applications}\ }\textbf {\bibinfo {volume} {168}},\ \bibinfo {pages}
  {561--580} (\bibinfo {year} {1990})}\BibitemShut {NoStop}%
\bibitem [{\citenamefont {{Ben Amar}}\ and\ \citenamefont
  {Goriely}(2005)}]{ben-amar-jmps-2005a}%
  \BibitemOpen
  \bibfield  {author} {\bibinfo {author} {\bibfnamefont {M.}~\bibnamefont {{Ben
  Amar}}}\ and\ \bibinfo {author} {\bibfnamefont {A.}~\bibnamefont {Goriely}},\
  }\bibfield  {title} {\enquote {\bibinfo {title} {Growth and instability in
  elastic tissues},}\ }\href {\doibase
  http://dx.doi.org/10.1016/j.jmps.2005.04.008} {\bibfield  {journal} {\bibinfo
   {journal} {Journal of the Mechanics and Physics of Solids}\ }\textbf
  {\bibinfo {volume} {53}},\ \bibinfo {pages} {2284--2319} (\bibinfo {year}
  {2005})}\BibitemShut {NoStop}%
\bibitem [{\citenamefont {Yin}\ \emph {et~al.}(2008)\citenamefont {Yin},
  \citenamefont {Cao}, \citenamefont {Li}, \citenamefont {Sheinman},\ and\
  \citenamefont {Chen}}]{yin-pnas-2008a}%
  \BibitemOpen
  \bibfield  {author} {\bibinfo {author} {\bibfnamefont {J.}~\bibnamefont
  {Yin}}, \bibinfo {author} {\bibfnamefont {Z.}~\bibnamefont {Cao}}, \bibinfo
  {author} {\bibfnamefont {C.}~\bibnamefont {Li}}, \bibinfo {author}
  {\bibfnamefont {I.}~\bibnamefont {Sheinman}}, \ and\ \bibinfo {author}
  {\bibfnamefont {X.}~\bibnamefont {Chen}},\ }\bibfield  {title} {\enquote
  {\bibinfo {title} {Stress-driven buckling patterns in spheroidal core/shell
  structures},}\ }\href {\doibase 10.1073/pnas.0810443105} {\bibfield
  {journal} {\bibinfo  {journal} {PNAS}\ }\textbf {\bibinfo {volume}
  {105}},\ \bibinfo {pages} {19132--19135} (\bibinfo {year}
  {2008})}\BibitemShut {NoStop}%
\bibitem [{\citenamefont {Ciarletta}\ \emph {et~al.}(2014)\citenamefont
  {Ciarletta}, \citenamefont {Balbi},\ and\ \citenamefont
  {Kuhl}}]{ciarletta-prl-2014a}%
  \BibitemOpen
  \bibfield  {author} {\bibinfo {author} {\bibfnamefont {P.}~\bibnamefont
  {Ciarletta}}, \bibinfo {author} {\bibfnamefont {V.}~\bibnamefont {Balbi}}, \
  and\ \bibinfo {author} {\bibfnamefont {E.}~\bibnamefont {Kuhl}},\ }\bibfield
  {title} {\enquote {\bibinfo {title} {Pattern selection in growing tubular
  tissues},}\ }\href {\doibase 10.1103/PhysRevLett.113.248101} {\bibfield
  {journal} {\bibinfo  {journal} {Physical Review Letters}\ }\textbf {\bibinfo
  {volume} {113}},\ \bibinfo {pages} {248101} (\bibinfo {year}
  {2014})}\BibitemShut {NoStop}%
\bibitem [{\citenamefont {Tallinen}\ \emph {et~al.}(2014)\citenamefont
  {Tallinen}, \citenamefont {Chung}, \citenamefont {Biggins},\ and\
  \citenamefont {Mahadevan}}]{tallinen-pnas-2014a}%
  \BibitemOpen
  \bibfield  {author} {\bibinfo {author} {\bibfnamefont {T.}~\bibnamefont
  {Tallinen}}, \bibinfo {author} {\bibfnamefont {J.~Y.}\ \bibnamefont {Chung}},
  \bibinfo {author} {\bibfnamefont {J.~S.}\ \bibnamefont {Biggins}}, \ and\
  \bibinfo {author} {\bibfnamefont {L.}~\bibnamefont {Mahadevan}},\ }\bibfield
  {title} {\enquote {\bibinfo {title} {Gyrification from constrained cortical
  expansion},}\ }\href {\doibase 10.1073/pnas.1406015111} {\bibfield  {journal}
  {\bibinfo  {journal} {PNAS}\ }\textbf {\bibinfo {volume} {111}},\ \bibinfo
  {pages} {12667--12672} (\bibinfo {year} {2014})}\BibitemShut {NoStop}%
\bibitem [{\citenamefont {Li}\ \emph {et~al.}(2011)\citenamefont {Li},
  \citenamefont {Jia}, \citenamefont {Cao}, \citenamefont {Feng},\ and\
  \citenamefont {Gao}}]{li-prl-2011a}%
  \BibitemOpen
  \bibfield  {author} {\bibinfo {author} {\bibfnamefont {B.}~\bibnamefont
  {Li}}, \bibinfo {author} {\bibfnamefont {F.}~\bibnamefont {Jia}}, \bibinfo
  {author} {\bibfnamefont {Y.-P.}\ \bibnamefont {Cao}}, \bibinfo {author}
  {\bibfnamefont {X.-Q.}\ \bibnamefont {Feng}}, \ and\ \bibinfo {author}
  {\bibfnamefont {H.}~\bibnamefont {Gao}},\ }\bibfield  {title} {\enquote
  {\bibinfo {title} {Surface wrinkling patterns on a core-shell soft sphere},}\
  }\href {\doibase 10.1103/PhysRevLett.106.234301} {\bibfield  {journal}
  {\bibinfo  {journal} {Physical Review Letters}\ }\textbf {\bibinfo {volume}
  {106}},\ \bibinfo {pages} {234301} (\bibinfo {year} {2011})}\BibitemShut
  {NoStop}%
\bibitem [{\citenamefont {Ciarletta}(2013)}]{ciarletta-prl-2013a}%
  \BibitemOpen
  \bibfield  {author} {\bibinfo {author} {\bibfnamefont {P.}~\bibnamefont
  {Ciarletta}},\ }\bibfield  {title} {\enquote {\bibinfo {title} {Buckling
  instability in growing tumor spheroids},}\ }\href {\doibase
  10.1103/PhysRevLett.110.158102} {\bibfield  {journal} {\bibinfo  {journal}
  {Physical Review Letters}\ }\textbf {\bibinfo {volume} {110}},\ \bibinfo
  {pages} {158102} (\bibinfo {year} {2013})}\BibitemShut {NoStop}%
\bibitem [{\citenamefont {Matsuo}\ and\ \citenamefont
  {Tanaka}(1992)}]{matsuo-nature-1992}%
  \BibitemOpen
  \bibfield  {author} {\bibinfo {author} {\bibfnamefont {E.~S.}\ \bibnamefont
  {Matsuo}}\ and\ \bibinfo {author} {\bibfnamefont {T.}~\bibnamefont
  {Tanaka}},\ }\bibfield  {title} {\enquote {\bibinfo {title} {Patterns in
  shrinking gels},}\ }\href {\doibase 10.1038/358482a0} {\bibfield  {journal}
  {\bibinfo  {journal} {Nature}\ }\textbf {\bibinfo {volume} {358}},\ \bibinfo
  {pages} {482--485} (\bibinfo {year} {1992})}\BibitemShut {NoStop}%
\bibitem [{\citenamefont {Boudaoud}\ and\ \citenamefont
  {Cha{\"{i}}eb}(2003)}]{boudaoud-pre-2003}%
  \BibitemOpen
  \bibfield  {author} {\bibinfo {author} {\bibfnamefont {A.}~\bibnamefont
  {Boudaoud}}\ and\ \bibinfo {author} {\bibfnamefont {S.}~\bibnamefont
  {Cha{\"{i}}eb}},\ }\bibfield  {title} {\enquote {\bibinfo {title} {Mechanical
  phase diagram of shrinking cylindrical gels},}\ }\href {\doibase
  10.1103/PhysRevE.68.021801} {\bibfield  {journal} {\bibinfo  {journal}
  {Phys. Rev. E}\ }\textbf {\bibinfo {volume} {68}},\ \bibinfo {pages}
  {021801} (\bibinfo {year} {2003})}\BibitemShut {NoStop}%
\bibitem [{\citenamefont {Dufresne}\ \emph {et~al.}(2006)\citenamefont
  {Dufresne}, \citenamefont {Stark}, \citenamefont {Greenblatt}, \citenamefont
  {Cheng}, \citenamefont {Hutchinson}, \citenamefont {Mahadevan},\ and\
  \citenamefont {Weitz}}]{dufresne-langmuir-2006a}%
  \BibitemOpen
  \bibfield  {author} {\bibinfo {author} {\bibfnamefont {E.~R.}\ \bibnamefont
  {Dufresne}}, \bibinfo {author} {\bibfnamefont {D.~J.}\ \bibnamefont {Stark}},
  \bibinfo {author} {\bibfnamefont {N.~A.}\ \bibnamefont {Greenblatt}},
  \bibinfo {author} {\bibfnamefont {J.~X.}\ \bibnamefont {Cheng}}, \bibinfo
  {author} {\bibfnamefont {J.~W.}\ \bibnamefont {Hutchinson}}, \bibinfo
  {author} {\bibfnamefont {L.}~\bibnamefont {Mahadevan}}, \ and\ \bibinfo
  {author} {\bibfnamefont {D.~A.}\ \bibnamefont {Weitz}},\ }\bibfield  {title}
  {\enquote {\bibinfo {title} {Dynamics of fracture in drying suspensions},}\
  }\href {\doibase 10.1021/la061251+} {\bibfield  {journal} {\bibinfo
  {journal} {Langmuir}\ }\textbf {\bibinfo {volume} {22}},\ \bibinfo {pages}
  {7144--7147} (\bibinfo {year} {2006})}\BibitemShut {NoStop}%
\bibitem [{\citenamefont {Goehring}\ \emph {et~al.}(2009)\citenamefont
  {Goehring}, \citenamefont {Mahadevan},\ and\ \citenamefont
  {Morris}}]{goehring-pnas-2009a}%
  \BibitemOpen
  \bibfield  {author} {\bibinfo {author} {\bibfnamefont {L.}~\bibnamefont
  {Goehring}}, \bibinfo {author} {\bibfnamefont {L.}~\bibnamefont {Mahadevan}},
  \ and\ \bibinfo {author} {\bibfnamefont {S.~W.}\ \bibnamefont {Morris}},\
  }\bibfield  {title} {\enquote {\bibinfo {title} {Nonequilibrium scale
  selection mechanism for columnar jointing},}\ }\href {\doibase
  10.1073/pnas.0805132106} {\bibfield  {journal} {\bibinfo  {journal}
  {PNAS}\ }\textbf {\bibinfo {volume} {106}},\ \bibinfo {pages} {387--392}
  (\bibinfo {year} {2009})}\BibitemShut {NoStop}%
\bibitem [{\citenamefont {Gong}\ \emph {et~al.}(2003)\citenamefont {Gong},
  \citenamefont {Katsuyama}, \citenamefont {Kurokawa},\ and\ \citenamefont
  {Osada}}]{gong-am-2003a}%
  \BibitemOpen
  \bibfield  {author} {\bibinfo {author} {\bibfnamefont {J.P.}\ \bibnamefont
  {Gong}}, \bibinfo {author} {\bibfnamefont {Y.}~\bibnamefont {Katsuyama}},
  \bibinfo {author} {\bibfnamefont {T.}~\bibnamefont {Kurokawa}}, \ and\
  \bibinfo {author} {\bibfnamefont {Y.}~\bibnamefont {Osada}},\ }\bibfield
  {title} {\enquote {\bibinfo {title} {Double-network hydrogels with extremely
  high mechanical strength},}\ }\href {\doibase 10.1002/adma.200304907}
  {\bibfield  {journal} {\bibinfo  {journal} {Advanced Materials}\ }\textbf
  {\bibinfo {volume} {15}},\ \bibinfo {pages} {1155--1158} (\bibinfo {year}
  {2003})}\BibitemShut {NoStop}%
\end{thebibliography}
%

\clearpage
\appendix
\section{Swelling in an Eulerian frame}
\label{app:Eulerian}

In an Eulerian frame, it is natural to work with so-called \textit{true} quantities, which measure the current stresses, fluxes, \textit{etc.} acting on or through the current (deformed) areas or volumes. For example, the true porosity $\phi_f$ measures the current fluid volume per unit current total volume. The solid displacement field is
\begin{equation}
    \mathbf{u}_s=\mathbf{x}-\mathbf{X}(\mathbf{x},t),
\end{equation}
where $\mathbf{x}$ is the Eulerian (spatial) coordinate and $\mathbf{X}(\mathbf{x},t)$ is the reference position of the material that is currently at position $\mathbf{x}$. We define the deformation gradient tensor $\mathbf{F}$ through its inverse,
\begin{equation}
    \mathbf{F}^{-1} =\boldsymbol{\nabla}\mathbf{X} =\mathbf{I}-\boldsymbol{\nabla}\mathbf{u}_s,
\end{equation}
where $\mathbf{I}$ is the identity tensor. The porosity is related to the Jacobian determinant $J$ via
\begin{equation}\label{eq:phi_to_J}
    J = \det{\mathbf{F}}=\frac{1}{1-\phi_f},
\end{equation}
where we assume that the fluid and solid constituents are individually incompressible and that the reference state is relaxed and dry ($\mathbf{u}_s=0\to\boldsymbol{\sigma}^\prime =\boldsymbol{0}\,,\,\phi_f=0$).
Continuity requires that
\begin{subequations}
    \begin{align}
        \frac{\partial{\phi_f}}{\partial{t}} &+\boldsymbol{\nabla}\cdot\left(\phi_f{}\mathbf{v}_f\right)=0\quad{}\mathrm{and} \label{eq:phi_law_vf} \\
    \frac{\partial{\phi_s}}{\partial{t}} &+\boldsymbol{\nabla}\cdot\left(\phi_s{}\mathbf{v}_s\right)=0,
    \end{align}
\end{subequations}
where $\mathbf{v}_f$ and $\mathbf{v}_s$ are the fluid and solid velocities and the true flux of fluid through the solid skeleton is (see Appendix~\ref{app:kinetics})
\begin{equation}
    \mathbf{w}_f=\phi_f\left(\mathbf{v}_f-\mathbf{v}_s\right) =-\frac{k(\phi_f)}{\eta\Omega_f}\boldsymbol{\nabla}\mu_f.
\end{equation}
In the absence of body forces, mechanical equilibrium requires that
\begin{equation}
    \boldsymbol{\nabla}\cdot\boldsymbol{\sigma}=0,
\end{equation}
where the true total stress $\boldsymbol{\sigma}$ is related to the true effective stress $\boldsymbol{\sigma}^\prime$ and the pore pressure $p$ via
\begin{equation}
    \boldsymbol{\sigma}=\boldsymbol{\sigma}^\prime-p\mathbf{I}.
\end{equation}
Finally, the chemical potential $\mu_f$ is given by
\begin{equation}
    \frac{\mu_f}{\Omega_f}=p-\Pi
\end{equation}
and the general expression for the Gaussian-chain constitutive law is \citep{cai-epl-2012a}
\begin{equation}
    J\boldsymbol{\sigma}^\prime =\frac{k_BT}{\Omega_p}\left(\mathbf{F}\mathbf{F}^\mathsf{T}-\mathbf{I}\right).
\end{equation}

\section{Swelling in a Lagrangian frame}
\label{app:Lagrangian}

In a Lagrangian frame, it is natural to work with so-called \textit{nominal} quantities, which measure the current stresses, fluxes, \textit{etc.} acting on or through the reference (relaxed) areas or volumes. For example, the nominal porosity $\Phi_f$ measures the current fluid volume per unit reference total volume, and is related to the true porosity via $\Phi_f=J\phi_f$. We denote the gradient and divergence operators in the Lagrangian coordinate system by $\mathrm{grad}(\cdot)$ and $\mathrm{div}(\cdot)$, respectively, to distinguish them from the corresponding operators in the Eulerian coordinate system.
The solid displacement field is
\begin{equation}
    \mathbf{U}_s=\mathbf{x}(\mathbf{X},t)-\mathbf{X},
\end{equation}
where $\mathbf{X}$ is the Lagrangian (material) coordinate and $\mathbf{x}(\mathbf{X},t)$ is the current position of the material associated with reference position $\mathbf{X}$. The deformation gradient tensor is then
\begin{equation}
    \mathbf{F} =\mathrm{grad}(\mathbf{x}) =\mathbf{I}+\mathrm{grad}(\mathbf{U}_s).
\end{equation}
The nominal porosity is related to the Jacobian determinant by
\begin{equation}
    J = \det{\mathbf{F}}=1+\Phi_f.
\end{equation}
Continuity requires that
\begin{equation}
    \frac{\partial{\Phi_f}}{\partial{t}}+\mathrm{div}\left(\mathbf{W}_f\right)=0,
\end{equation}
where $\mathbf{W}_f$ is the nominal flux of fluid through the solid skeleton,
\begin{equation}
    \mathbf{W}_f =-J\mathbf{F}^{-1}\mathbf{F}^{-\mathsf{T}}\,\frac{k(\phi_f)}{\eta\Omega_f}\,\mathrm{grad}(\mu_f).
\end{equation}
Mechanical equilibrium requires that
\begin{equation}
    \mathrm{div}(\mathbf{s})=0
\end{equation}
where the nominal total stress $\mathbf{s}$ is related to the nominal effective stress $\mathbf{s}^\prime$ and the pore pressure $p$ via
\begin{equation}
    \mathbf{s}=\mathbf{s}^\prime-J\mathbf{F}^{-\mathsf{T}}p,
\end{equation}
where
\begin{equation}
    \mathbf{s}=J\boldsymbol{\sigma}\mathbf{F}^{-\mathsf{T}} \quad\mathrm{and}\quad \mathbf{s}^\prime=J\boldsymbol{\sigma}^\prime\mathbf{F}^{-\mathsf{T}}.
\end{equation}
The chemical potential is again given by
\begin{equation}
    \frac{\mu_f}{\Omega_f}=p-\Pi.
\end{equation}

\clearpage
\section{Composition, porosity, \\ and free energy of mixing}
\label{app:concentration}

The free energy of mixing $\mathcal{F}_\mathrm{mix}$ is typically taken to be a function of the true number density of water molecules $n_f$, or that of polymer molecules $n_p$ (number of molecules per unit volume of mixture). These densities can then be related to the porosity $\phi_f$, which measures the volume of fluid per unit volume of mixture,
\begin{equation}
    \phi_f=\Omega_fn_f=1-\Omega_pn_p
\end{equation}
where $\Omega_f$ and $\Omega_p$ are the volume per molecule of water and polymer, respectively, \textit{in their unmixed states}. It is typically assumed that these volumes are unchanged upon mixing and deformation. Recalling that $\phi_f$ is related to the Jacobian determinant $J$ via Eq.~\eqref{eq:phi_to_J}, we have that
\begin{equation}
    J=\frac{1}{1-\phi_f}=\frac{1}{1-\Omega_fn_f}=\frac{1}{\Omega_pn_p}.
\end{equation}
The local chemical composition is therefore uniquely characterized by $\phi_f$ or $J$. Note that the nominal number densities $N_f$ and $N_p$ (number of molecules per unit reference volume of dry polymer) are related to the true number densities via $N_f=Jn_f$ and $N_p=Jn_p$.

\section{Flory-Huggins free energy}
\label{app:FloryHuggins}

For a polymer solution, the classical Flory-Huggins free energy of mixing per unit reference volume can be written~\citep{flory-jchemphys-1942, huggins-jphyschem-1942}
\begin{equation}
    \mathcal{F}_\mathrm{mix} =J\,\frac{k_BT}{\Omega_f}\bigg[ \phi_f\ln\phi_f +\frac{1}{\alpha}\phi_s\ln\phi_s+\chi\phi_f\phi_s\bigg],
\end{equation}
where $\phi_s\equiv{}1-\phi_f$ is the true solid fraction. The prefactor $J$ converts the free energy per unit current volume to the free energy per unit reference volume. The first two terms in square brackets reflect the entropy of mixing, where $\alpha$ is a measure of the volume per polymer chain relative to the volume per fluid molecule in the mixture. The third term in square brackets reflects the enthalpy of mixing, where $\chi$ is the dimensionless interaction parameter. It is straightforward to rewrite this expression in terms of $J$.

Although the two parameters $\alpha$ and $\chi$ have meaningful physical interpretations, these are typically used as fitting parameters to account for the various approximations embedded in this theory~\citep[\textit{e.g.},][]{cai-epl-2012a, li-softmatter-2012a, engelsberg-pre-2013a}.

\section{Transport law}
\label{app:kinetics}

The true flux of fluid through the solid skeleton is often modelled as a diffusive process driven by gradients in chemical potential,
\begin{equation}
    \mathbf{w}_f =\phi_f(\mathbf{v}_f-\mathbf{v}_s)=-\frac{D(\phi_f)}{k_BT\Omega_f}\boldsymbol{\nabla}\mu_f,
\end{equation}
where $k_B$ is the Boltzmann constant, $T$ is the absolute temperature, and $D(\phi_f)$ is the effective diffusion coefficient. The effective diffusion coefficient is, in general, a function of the local composition, as measured by $\phi_f$. From the perspective of chemical kinetics, this can capture linear diffusion (Fick's law) by taking $D(\phi_f)=D_0$, where $D_0$ is a constant, or type-II diffusion with a flux proportional to the local volume fraction of fluid by taking $D(\phi_f)=D_0\phi_f$. From the perspective of flow through porous media, this can be reinterpreted as Darcy's law by taking $D(\phi_f)=(k_BT/\eta)k(\phi_f)$, where $\eta$ is the dynamic viscosity of the fluid and $k(\phi_f)$ is the permeability of the solid skeleton. Fick's law and Darcy's law provide equivalent descriptions of water transport within the gel~\citep[see][]{peppin-physfluids-2005a}.

The form of the permeability function should incorporate the geometry of the polymer network, with the most important feature being that the permeability should increase very strongly with increasing fluid content. For polymeric gels, the frictional drag $f$ between water and polymer is typically taken to be inversely proportional to the square of the characteristic mesh size $l$, or $f\sim{}l^{-2}$. The mesh size is itself related to the correlation length (distance between crosslinks), and can be taken to be proportional to $(1-\phi_f)^{-3/4}$~\citep{tokita-jcp-1991a}. This leads to $f\sim{}(1-\phi_f)^{3/2}$, and therefore to a permeability function $k(\phi_f)\sim{}\phi_ff^{-1}\sim{}\phi_f(1-\phi_f)^{-\beta}$ with $\beta=3/2$. This expression has subsequently been used in a variety of studies, some of which take $\beta$ as an empirical fitting parameter~\citep{engelsberg-pre-2013a, grattoni-jcis-2001a}.

Here, we simply take $\beta=3/2$ (cf., Eq.~18) and our modelling predictions ultimately agree very well with our experimental results for this value. Of course, the model itself is valid for any form of the permeability law (Eq.~19). The precise form is unlikely to change the qualitative features of swelling and drying, which is ultimately the focus of our study.

\section{Numerical integration}
\label{app:Numerics}

To formulate a finite-volume scheme, we first divide the interval $\tilde{r}=[0,\tilde{a}]$ into $N$ elements of equal size $\delta{\tilde{r}}=\tilde{a}/N$, where element $i$ has its center at $\tilde{r}_i=(i-1/2)\delta{\tilde{r}}$ and its left and right edges at $\tilde{r}_{i-1/2}=(i-1)\delta{\tilde{r}}$ and $\tilde{r}_{i+1/2}=i\delta{\tilde{r}}$, respectively. We then calculate
\begin{subequations}\label{eq:drdefs}
    \begin{align}
        \frac{\partial}{\partial{\tilde{t}}}\,\delta{\tilde{r}} &=\frac{1}{N}\,\frac{\mathrm{d}\tilde{a}}{\mathrm{d}\tilde{t}} =\frac{\delta{\tilde{r}}}{\tilde{a}}\,\frac{\mathrm{d}\tilde{a}}{\mathrm{d}\tilde{t}},
    \end{align}
and
    \begin{align}
        \frac{\partial}{\partial{\tilde{t}}}\,\tilde{r}_i &=(i-1/2)\,\frac{\mathrm{d}}{\mathrm{d}\tilde{t}}\,\delta{\tilde{r}} =\frac{\tilde{r}_i}{\tilde{a}}\,\frac{\mathrm{d}\tilde{a}}{\mathrm{d}\tilde{t}}.
    \end{align}
\end{subequations}
We then integrate the conservation law over element $i$,
\begin{equation}
    \begin{split}
        \int_{\tilde{r}_{i-1/2}}^{\tilde{r}_{i+1/2}}\,4\pi{}\tilde{r}^2\,\mathrm{d}\tilde{r}\,\bigg\{ \frac{\partial{\phi_f}}{\partial{\tilde{t}}}& \\
        -\frac{1}{\tilde{r}^2}\frac{\partial}{\partial{\tilde{r}}}\bigg[\tilde{r}^2&(1-\phi_f)\tilde{k}(\phi_f)\frac{\partial{\tilde{\mu}}}{\partial{\tilde{r}}}\bigg]\bigg\}=0.
    \end{split}
\end{equation}
After some algebra, and making use of Eqs.~\eqref{eq:drdefs} and the Leibnitz integral rule, we arrive at
\begin{equation}\label{eq:philawFV}
    \begin{split}
        \frac{4}{3}\pi&\left(\tilde{r}_{i+1/2}^3-\tilde{r}_{i-1/2}^3\right)\left(\frac{\partial{\phi_{f,i}}}{\partial{\tilde{t}}} +\frac{3\phi_{f,i}}{\tilde{a}}\,\frac{\mathrm{d}\tilde{a}}{\mathrm{d}\tilde{t}}\right) \\
        &-4\pi\left[\frac{\tilde{r}^3\phi_f}{\tilde{a}}\,\frac{\mathrm{d}\tilde{a}}{\mathrm{d}\tilde{t}} +\tilde{r}^2\,(1-\phi_f)\,\tilde{k}(\phi_f)\,\ \frac{\partial{\tilde{\mu}_f}}{\partial{\tilde{r}}}\right]\bigg|_{\tilde{r}_{i-1/2}}^{\tilde{r}_{i+1/2}}=0,
    \end{split}
\end{equation}
where $\phi_{f,i}$ is the mean porosity in element $i$. We then require boundary conditions at $\tilde{r}=0$ and $\tilde{r}=\tilde{a}$, for which it is useful to recall that the second term in square brackets is precisely equal to $-\tilde{r}^2\phi_f\tilde{v}_f$ (see Eq.~15a). At $\tilde{r}=0$, the entire quantity in square brackets must vanish. At $\tilde{r}=\tilde{a}$, the entire quantity is identically equal to $\tilde{a}^2\mathrm{d}\tilde{a}/\mathrm{d}t$.

At each time step, we calculate $u_s$ from $\phi_f$ via Eq.~(14). We then calculate $\lambda_r$, $\lambda_\theta$, and $J$ from $u_s$, then $\sigma_r^\prime$, $\sigma_\theta^\prime$, and $\Pi$ from the constitutive laws, and then $\partial{\tilde{\mu}_f}/\partial{\tilde{r}}$ from Eq.~(20). We finally use this to update the porosity according to Eq.~\eqref{eq:philawFV}.

\section{Equilibrium size in air}
\label{app:drysize}

\begin{figure}[t!]
    \centering
    \includegraphics[width=8.6cm]{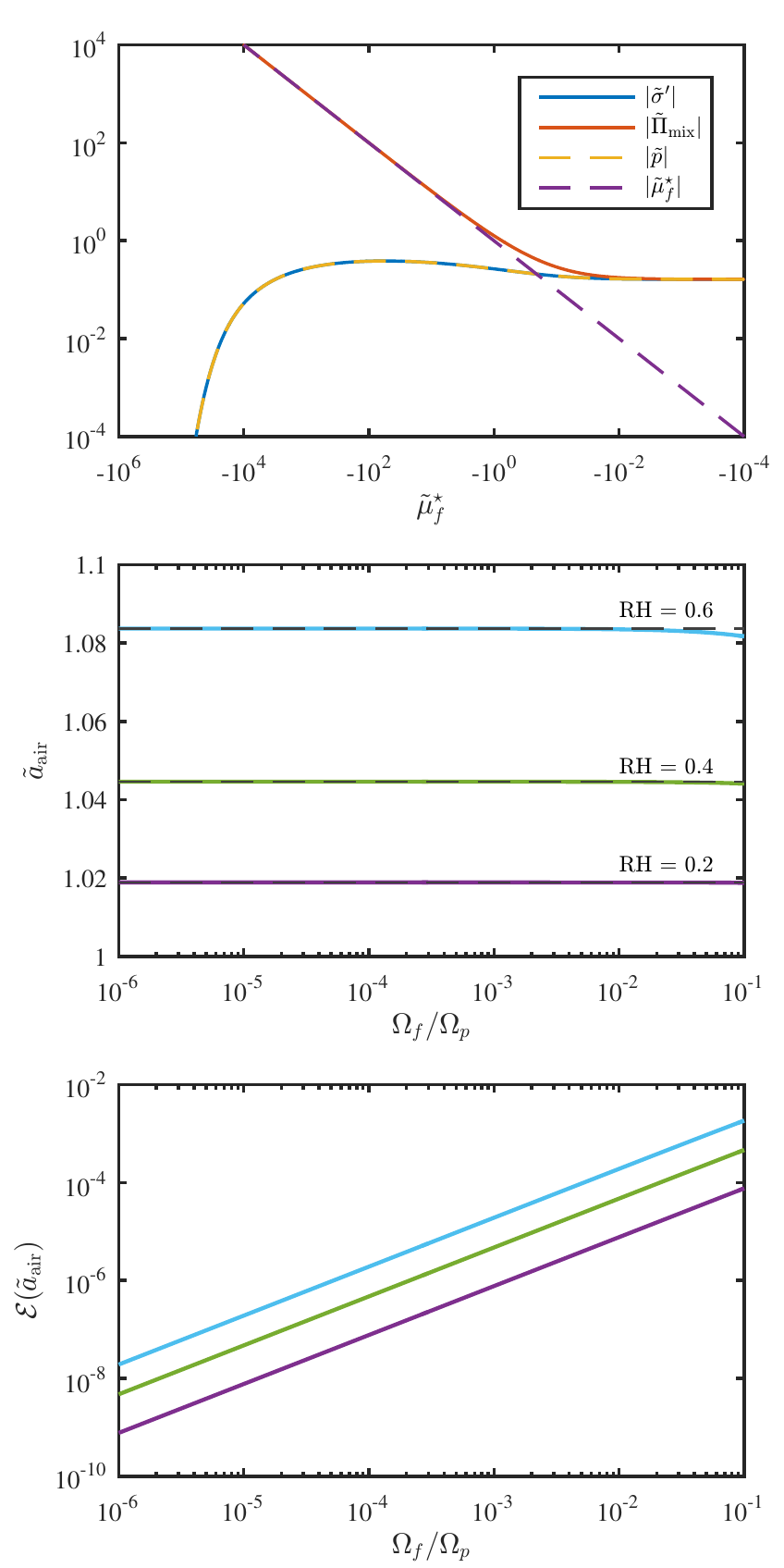}
    \caption{The size of a hydrogel sphere in air is effectively independent of $\Omega_f/\Omega_p$. Top:~Properties of the equilibrium state for a wide range of ambient conditions ($\tilde{\mu}_f^\star$) with material properties $\Omega_f/\Omega_p=1.28\times{}10^{-4}$, $\alpha=250$, and $\chi=0.4$. Middle:~Actual equilibrium size in air $\tilde{a}_\mathrm{air}$ as a function of $\Omega_f/\Omega_p$ for several values of $\mathrm{RH}$ (colors) compared with the value of $\tilde{a}_\mathrm{air}$ for the same $\mathrm{RH}$ for $\Omega_f/\Omega_p\to{}0$ (dashed gray). Bottom:~The relative error between the colored and gray curves from the middle figure. \label{fig:drysize} }
\end{figure}

The equilibrium size in air is effectively independent of $\Omega_f/\Omega_p$ because, for $\tilde{\mu}_f^\star$ less than about $-10^{2}$, the mechanical contributions to the equilibrium state ($\tilde{p}$ and $\tilde{\sigma}^\prime$) become negligible relative to the chemical contributions ($\tilde{\mu}_f^\star$ and $\tilde{\Pi}_\mathrm{mix}$) since the polymer chains are nearly relaxed (see the main text, after Eqs.~29). We plot the magnitudes of these contributions against $\tilde{\mu}_f^\star$ in Fig.~\ref{fig:drysize} (top). We confirm this in Fig.~\ref{fig:drysize} (middle and bottom) by plotting the equilibrium size against $\Omega_f/\Omega_p$ for several values of $\tilde{\mu}_f^\star$ ($\mathrm{RH}$) and comparing these with the dry size for $\Omega_f/\Omega_p\to{}0$.

\section{Compressive and tensile stresses during swelling}
\label{app:stressmap}

\begin{figure}[t!]
    \centering
    \includegraphics[width=8.6cm]{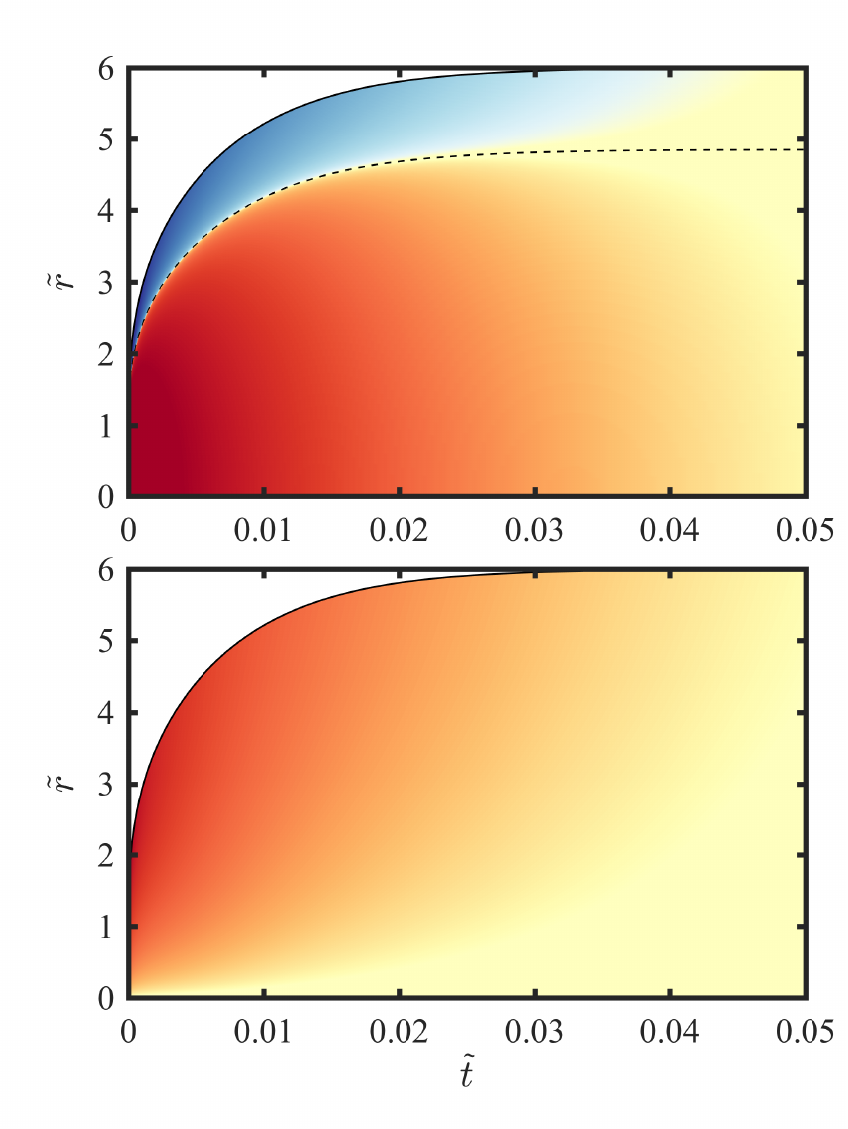}
    \caption{Space-time evolution of (top)~the mean total stress $\bar{\tilde{\sigma}}=(\tilde{\sigma}_r+\tilde{\sigma}_\theta)/2$ and (bottom)~the shear stress $\tilde{\tau}=|\tilde{\sigma}_r-\tilde{\sigma}_\theta|/2$. The colors show $\mathrm{sign}(\tilde{\sigma})\log|\tilde{\sigma}|$, where blue tones are compressive, red tones are tensile, and the dashed black line in the top panel indicates the contour of zero mean total stress. \label{fig:stressmap} }
\end{figure}

During swelling, the outer shell is in a strong and anisotropic state of compression while the inner core is in a more isotropic state of tension (Fig.~\ref{fig:stressmap}).

\section{Evidence of a \\ core-shell structure}
\label{app:coreshell}

\begin{figure}[h!]
   \centering
    \includegraphics[width=8.6cm]{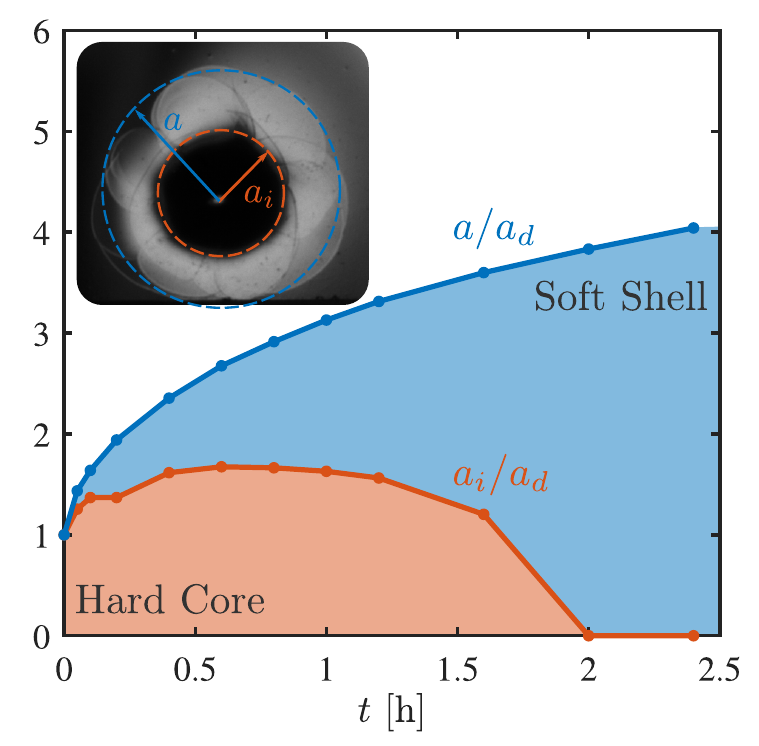}
    \caption{We image the swelling process using a shadowgraph technique, revealing two distinct regions in the internal structure: A dark, low-porosity core surrounded by a light, high-porosity shell (inset). Thresholding this image provides the time evolution of the outer radius of the core $a_i$ and of the sphere $a$, which together define the shell. We shade the core and shell regions in orange and blue, respectively. \label{fig:shadowgraph} }
\end{figure}

The porosity within the sphere becomes heterogeneous during swelling, developing a core-shell structure. 
Direct observation of the core-shell structure is complicated by the fact that the sphere is transparent, and the swollen region is almost entirely water. \citet{barros-softmatter-2012a} provided the first direct observation of this by imaging a swelling sphere using nuclear magnetic resonance (NMR). Here, we achieve a similar result with a shadowgraph technique~(Fig.~\ref{fig:shadowgraph}). We obtain images by collimating light from a powerful laser source ($1\,\mathrm{W}$, $532\,\mathrm{nm}$) via a ShadowStrobe lens (Dantec Dynamics). We identify the position of the core-shell interface via an intensity threshold and we plot the evolution of the core-shell structure in Fig.~\ref{fig:shadowgraph}. At early times, both core and shell grow as the sphere swells. Later, the core shrinks as water eventually imbibes into the core of the bead. The interface position detected through this method is qualitative since the relationship between light intensity and polymer density is unknown and likely nonlinear, but our findings are consistent with the predictions of our model.

\begin{figure}[h!]
    \centering
    \includegraphics[width=8.6cm]{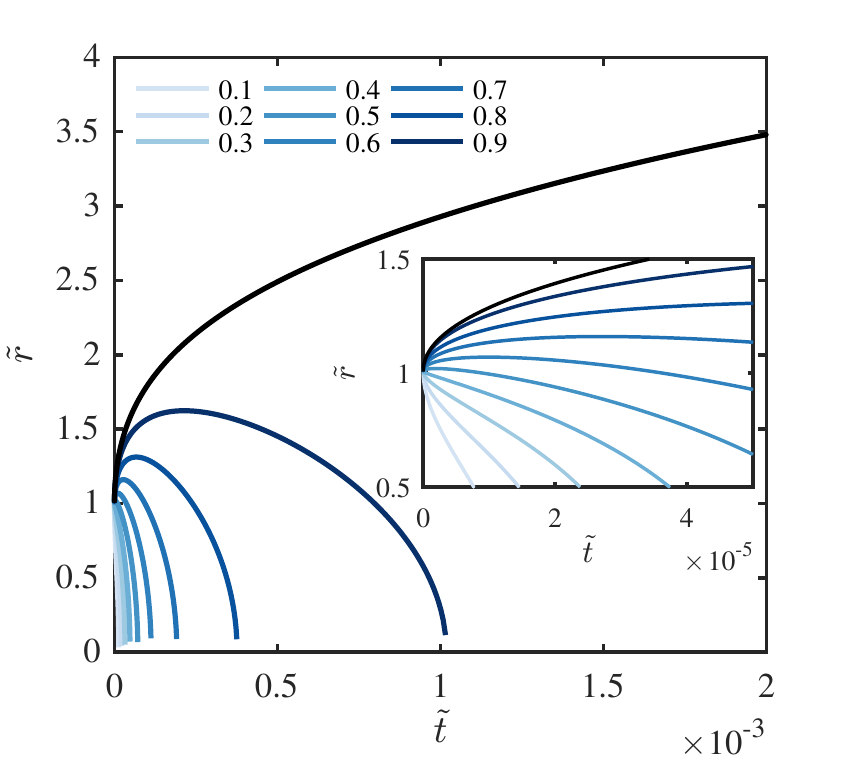}
    \caption{The positions of several isolines of porosity during swelling. The black line marks the outer radius of the sphere and the inset highlights the early-time evolution. \label{fig:porosity_isolines} }
\end{figure}

In contrast with our observations, the NMR experiments of \citet{barros-softmatter-2012a} and \citet{engelsberg-pre-2013a} suggest a strictly shrinking core. To reconcile this apparent disagreement, we plot in Fig.~\ref{fig:porosity_isolines} the predictions of our model for the location of several isolines of porosity against time. We find that, for porosities greater than about 0.5, the isolines initially advance and then retreat. For smaller porosities, the isolines strictly retreat. Assuming that the core revealed by both shadowgraph and NMR is roughly coincident with a certain porosity threshold, this then indicates that the apparent evolution of the core will depend on the threshold value associated with each technique. The qualitative agreement between the evolution of the core from our shadowgraph experiments (Fig.~\ref{fig:shadowgraph}) and the evolution of porosity isolines from the model for $\phi_f>0.5$ (Fig.~\ref{fig:porosity_isolines}) supports the kinetic predictions of the model and further underscores its usefulness for interpreting experimental results.

\begin{figure}[t]
    \centering
    \includegraphics[width=8.6cm]{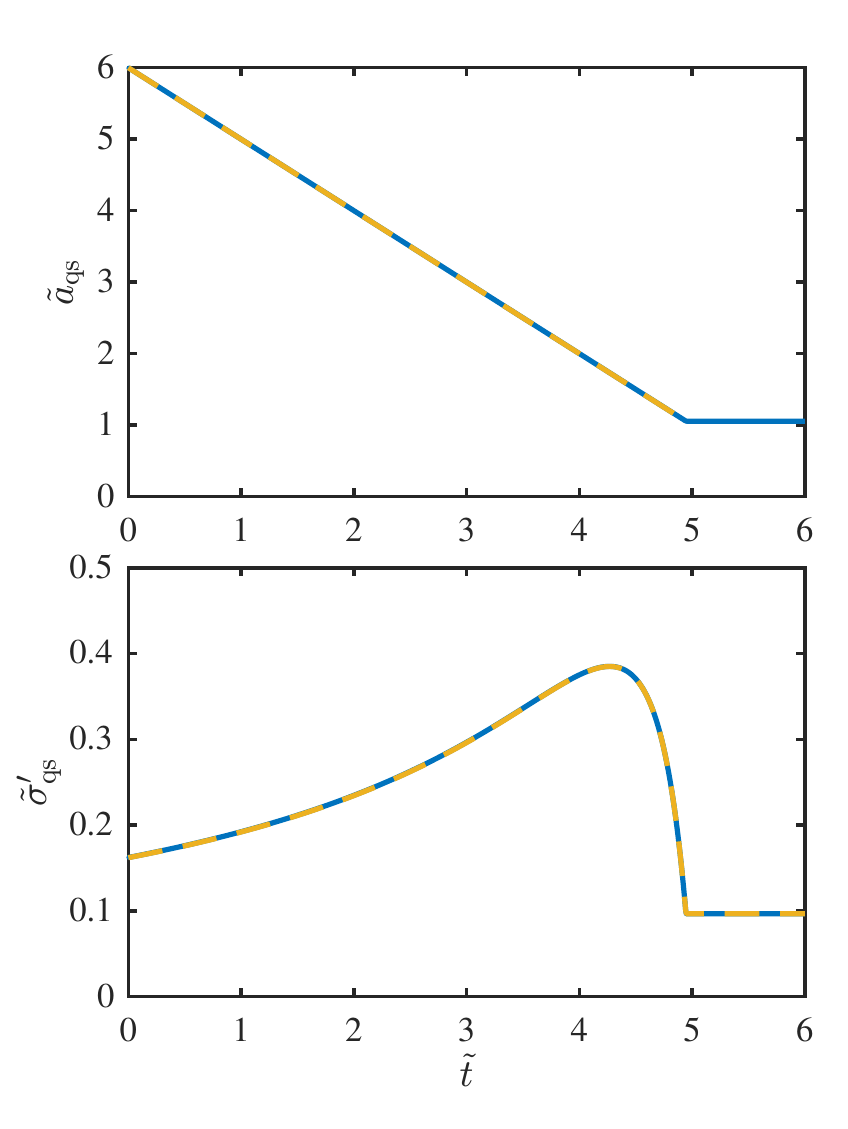}
    \caption{Evolution of the outer radius $\tilde{a}_\mathrm{qs}$ and the effective stress $\tilde{\sigma}_\mathrm{qs}^\prime$ during strongly limited drying from the full model (solid blue) and from the quasi-static model (dashed yellow). Parameters are the same as Fig.~\ref{fig:drying_model}, but with $\tilde{F}_d^\star=1$. \label{fig:drying_model_equilibrium} }
\end{figure}

\section{Quasi-static model}
\label{app:QSmodel}

When the flux of fluid out of the bead during drying is strongly limited (\textit{e.g.}, by evaporation), drying can be modeled as a quasi-static process in which the sphere is internally homogeneous. The same is true of flux-limited swelling. To develop a model for this, we first assume that the drying flux is controlled by the evaporation limit,
\begin{equation}
    \tilde{F}_{d,\mathrm{qs}} =-\frac{\mathrm{d}\tilde{a}_\mathrm{qs}}{\mathrm{d}\tilde{t}}=\tilde{F}_d^\star.
\end{equation}
This can be integrated to give
\begin{equation}
    \tilde{a}_\mathrm{qs}(\tilde{t})=\left\{\begin{array}{ll}
        \tilde{a}_0-\tilde{F}_d^\star\tilde{t} &\quad\mathrm{for}\quad \tilde{t}\leq\tilde{t}_\mathrm{eq} \\
        \tilde{a}_\mathrm{eq} &\quad\mathrm{for}\quad \tilde{t}>\tilde{t}_\mathrm{eq},
    \end{array}\right.
\end{equation}
where $\tilde{t}_\mathrm{eq}=(\tilde{a}_0-\tilde{a}_\mathrm{eq})/\tilde{F}_d^\star$ and $\tilde{a}_
\mathrm{eq}$ is the final equilibrium size for the desired value of $\tilde{\mu}_f^\star$. We can then calculate all other quantities from Eqs.~(29) by replacing $\tilde{a}_\mathrm{eq}$ with $\tilde{a}_\mathrm{qs}(\tilde{t})$. In particular, the uniform and isotropic effective stresses are given by
\begin{equation}
    \tilde{\sigma}_\mathrm{qs}^\prime(\tilde{t}) =\tilde{\sigma}_{r,\mathrm{qs}}^\prime(\tilde{t}) =\tilde{\sigma}_{\theta,\mathrm{qs}}^\prime(\tilde{t}) =[\tilde{a}_\mathrm{qs}(t)^2-1]/\tilde{a}_\mathrm{qs}(t)^3.
\end{equation}
It is then trivial to show that the effective stress has a tensile maximum of $\mathrm{max}_{\,t}\{\tilde{\sigma}_\mathrm{qs}^\prime\} = 2/(3\sqrt{3})\approx{}0.3849$ at $\tilde{a}_\mathrm{qs}=\sqrt{3}$.
We plot $\tilde{a}_\mathrm{qs}$ and $\tilde{\sigma}_\mathrm{qs}^\prime$ against $\tilde{t}$ in Fig.~\ref{fig:drying_model_equilibrium}.

\section{Time-reversibility of small deformations}
\label{app:smallswelling}

\begin{figure*}
    \centering
    \includegraphics[width=17.2cm]{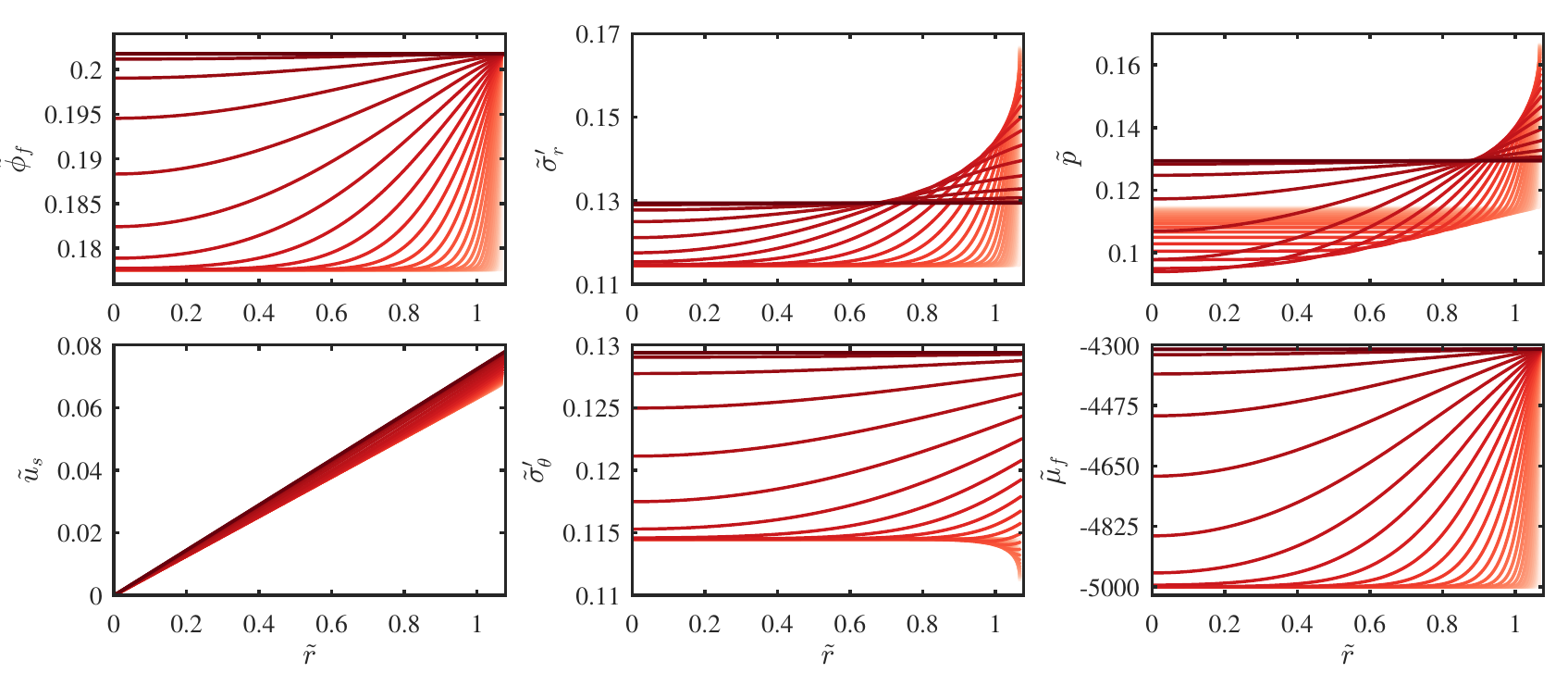}
    \caption{Free swelling for a small change in size, from $\tilde{a}_0=1.067$ to $\tilde{a}_\mathrm{eq}=1.078$ ($\tilde{\mu}_{f,0}^\star=-5\times{}10^3$ to $\tilde{\mu}_f^\star=-4.3\times{}10^{3}$). Same material properties as Fig.~\ref{fig:drying_model}. \label{fig:swelling_model_small} }
\end{figure*}

\begin{figure*}
    \centering
    \includegraphics[width=17.2cm]{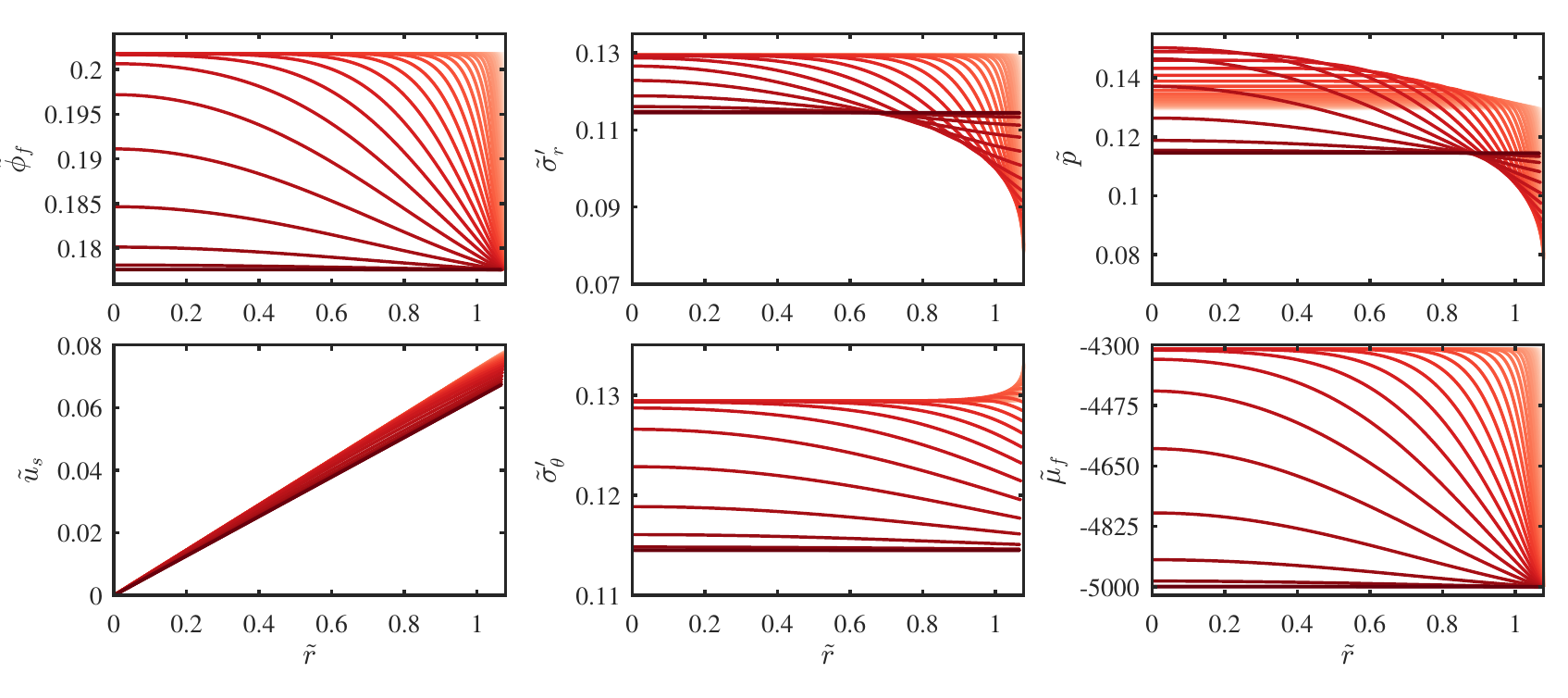}
    \caption{Free drying for a small change in size. Same material properties as Fig.~\ref{fig:swelling_model_small}, but with initial and final states reversed. \label{fig:drying_model_small} }
\end{figure*}

For small changes in size, swelling and drying are essentially mirror images of each other because the strong nonlinearity of large deformations is absent. We show swelling in Fig.~\ref{fig:swelling_model_small} and drying in Fig.~\ref{fig:drying_model_small}.


\section{Evaporation-limited drying}
\label{app:drying_constrained}

\begin{figure*}
    \centering
    \includegraphics[width=17.2cm]{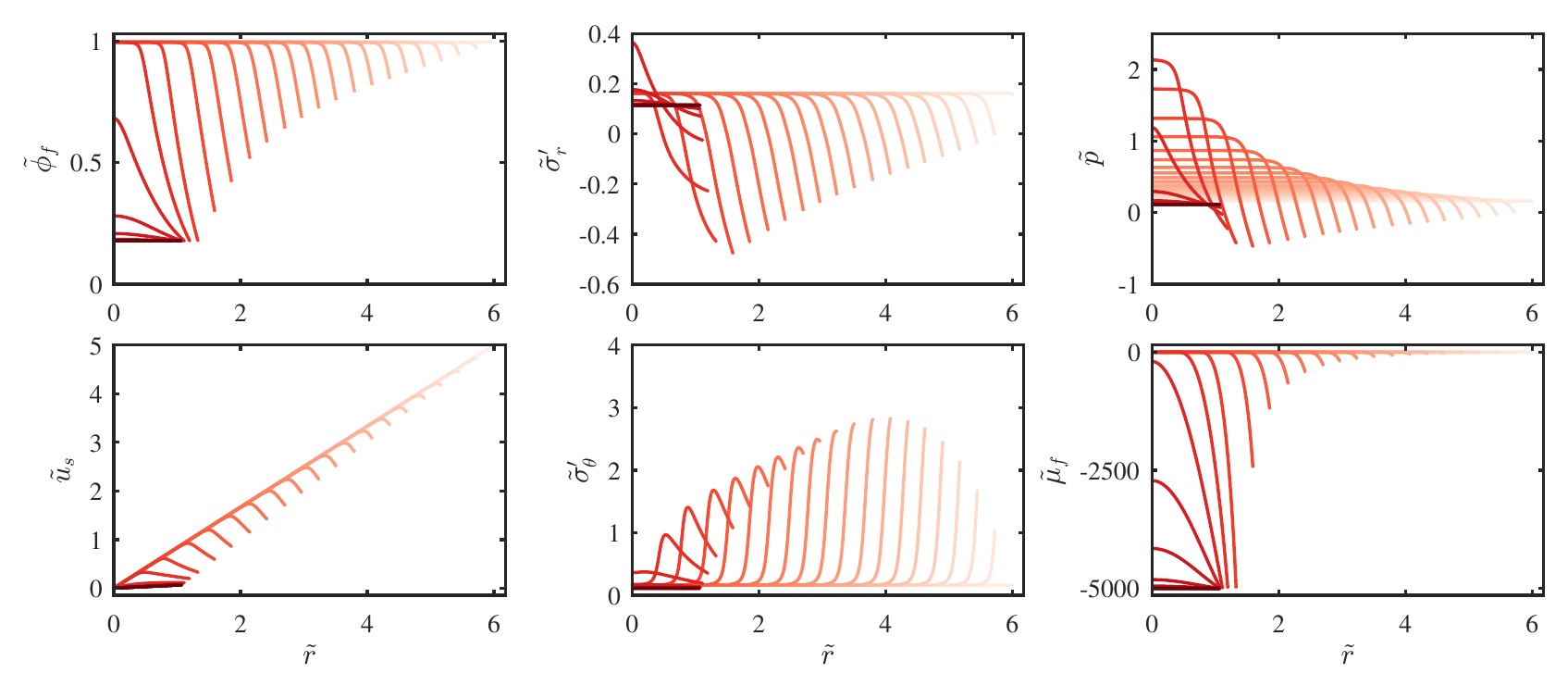}
    \caption{Evaporation-limited drying for $\tilde{F}_d^\star=10^{4}$. Same material properties and conditions as Fig.~\ref{fig:drying_model}. \label{fig_drying_model_constrained} }
\end{figure*}

We plot in Fig.~\ref{fig_drying_model_constrained} the evolution of a sphere during evaporation-limited drying (\textit{cf}., Fig.~\ref{fig:drying_model}). We enforce the limit $\tilde{F}_d(t)\leq\tilde{F}_d^\star$ by calculating, at every time, a new ambient value $\tilde{\mu}_{f,d}^\star(t)$ for which $\tilde{F}_d(t)=\tilde{F}_d^\star$ when $\tilde{\mu}_f(\tilde{a},\tilde{t})=\tilde{\mu}_{f,d}^\star(t)$. We then impose $\tilde{\mu}_f(\tilde{a},\tilde{t})=\max\{\tilde{\mu}_f^\star,\tilde{\mu}_{f,d}^\star(t)\}$ so that this constraint can only slow the drying process. As a result, $\tilde{\mu}_{f,d}^\star(t)$ evolves gradually toward the true ambient value $\tilde{\mu}_f^\star$ rather than adopting it immediately, as it would in free drying. This leads to much lower azimuthal effective stresses and much weaker gradients in porosity near the outer boundary.

\section{Drying experiments: Free drying}
\label{app:freedrying}

\begin{figure*}
    \centering
    \includegraphics[width=17.2cm]{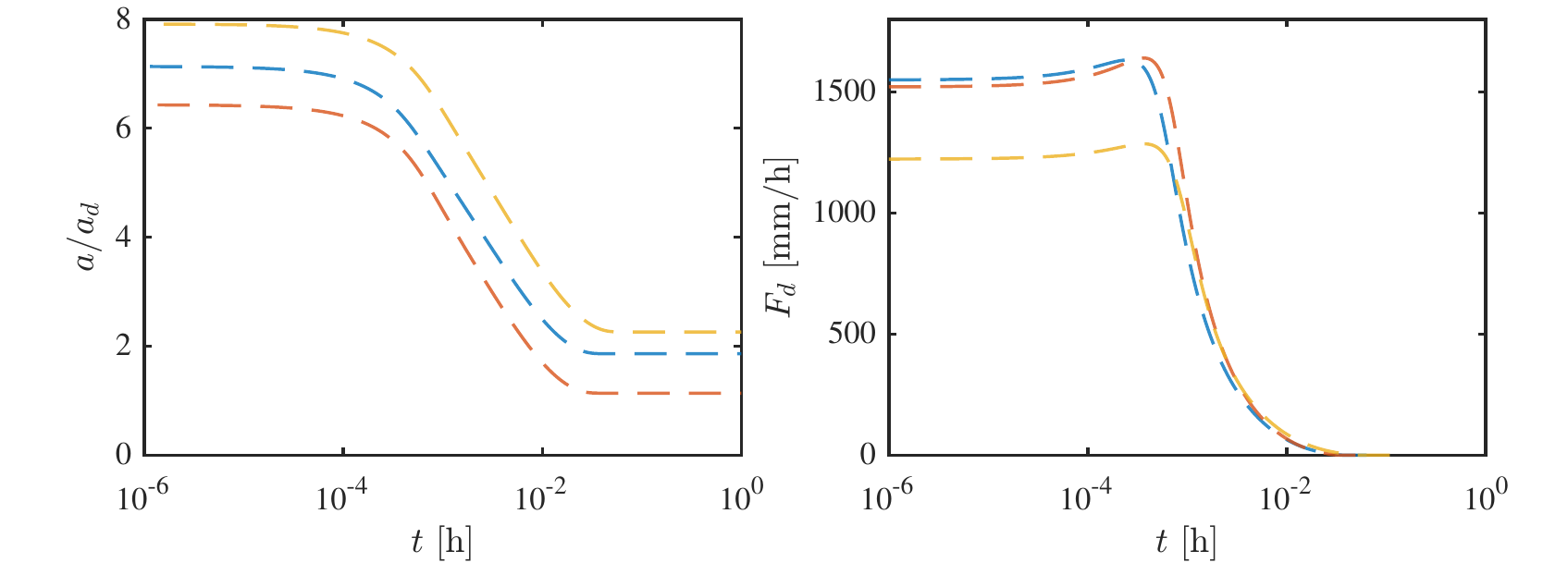}
    \caption{Free drying for the same parameters and conditions as Fig.~\ref{fig:drying_experiment}, but taking $F_d^\star\to\infty$. \label{fig:drying_experiment_free} }
\end{figure*}

To illustrate that our drying experiments are not in a state of free drying, we plot in Fig.~\ref{fig:drying_experiment_free} the time evolution of $a/a_d$ and $F_d$ for the same parameters as Fig.~\ref{fig:drying_experiment}, but taking $F_d^\star\to\infty$ (\textit{i.e.}, free drying). Note the very short time scale and the very large drying fluxes compared to the data.

\end{document}